\documentclass[aps,prx,reprint,twocolumn,amsmath,amssymb,floatfix,nofootinbib,superscriptaddress]{revtex4-2}
\usepackage{amsmath,amssymb,amsthm,booktabs,graphicx,hyperref}
\usepackage{tikz,pgfplots}
\pgfplotsset{compat=1.18}
\usetikzlibrary{arrows.meta,calc,positioning,decorations.pathreplacing,shapes.geometric}
\DeclareMathVersion{figsans}
\SetSymbolFont{operators}{figsans}{OT1}{phv}{m}{n}
\SetSymbolFont{letters}{figsans}{OML}{ntxsfmi}{m}{it}
\SetSymbolFont{symbols}{figsans}{OMS}{cmsy}{m}{n}
\SetMathAlphabet{\mathrm}{figsans}{OT1}{phv}{m}{n}
\SetMathAlphabet{\mathbf}{figsans}{OT1}{phv}{b}{n}
\hypersetup{hidelinks}
\newtheorem{theorem}{Theorem}
\newtheorem{lemma}[theorem]{Lemma}
\newtheorem{proposition}[theorem]{Proposition}
\newtheorem{corollary}[theorem]{Corollary}
\theoremstyle{definition}
\newtheorem{definition}[theorem]{Definition}
\theoremstyle{remark}

\newcommand{\F}{\mathbb F_2}
\newcommand{\CCZ}{\mathrm{CCZ}}

\newcommand{\wt}{\operatorname{wt}}
\newcommand{\fs}{\mathbb V}

\begin{document}

\title{The limits of magic-state cultivation}
\author{Lloren\c{c} Balada Gaggioli}
\affiliation{Czech Technical University in Prague, Prague, Czech Republic}
\affiliation{LAAS-CNRS, Universit\'e de Toulouse, France}
\author{Jakub Mare\v{c}ek}
\affiliation{Czech Technical University in Prague, Prague, Czech Republic}
\date{\today}
\begin{abstract}
Fault tolerance is understood through Pauli algebra, which fails at non-Clifford gates. Circuits such as magic-state cultivation, which checks an encoded T state with physical T gates, have therefore lacked an exact theory of which faults are harmful. We develop such a theory and show that, for any circuit that prepares a magic state with non-Clifford gates and postselection, one stabilizer code on the non-Clifford gates decides which faults are harmful. This ties cultivation to distillation, bounds its T-count from below and gives an exact test. For a single check on an ideally encoded state, the fault distance reduces to Pauli algebra on the code. As a use case, we analyze the S gates that have stood in for T gates in the design of cultivation. We show that this proxy can only overstate the fault distance of such a check, and that the checks in use, on color and surface codes, tolerate at most half as many faults as their code allows. The theory also yields nearest-neighbor checks that keep the full distance, certified up to code distance nine.
\end{abstract}
\maketitle

\begin{figure*}[t]
\centering
\definecolor{tolblue}{cmyk}{1,0.36,0,0.30}      
\definecolor{harmver}{cmyk}{0,0.56,1,0.165}     
\definecolor{skyblue}{cmyk}{0.63,0.23,0,0.09}   
\definecolor{rejgray}{cmyk}{0,0,0,0.62}         
\definecolor{inkgray}{cmyk}{0,0,0,0.75}         
\providecommand{\figPi}{\text{\fontencoding{OT1}\fontfamily{cmss}\selectfont\char5}}
\begin{tikzpicture}[
  font=\fontsize{8}{9.6}\fontfamily{phv}\selectfont\mathversion{figsans},
  line width=0.5pt,
  >={Stealth[length=1.7mm,width=1.3mm]},
  letter/.style={font=\fontsize{8}{9.6}\fontfamily{phv}\fontseries{b}\selectfont, anchor=north west, inner sep=0pt},
  small/.style={font=\fontsize{7}{8.4}\fontfamily{phv}\selectfont\mathversion{figsans}},
  gate/.style={draw, rounded corners=1pt, fill=black!8},
  tgate/.style={draw=tolblue, rounded corners=1pt, fill=tolblue!12, minimum width=0.36cm, minimum height=0.24cm, inner sep=0pt, font=\fontsize{7}{8}\fontfamily{phv}\selectfont},
  code/.style={draw=skyblue, rounded corners=1pt, fill=skyblue!18},
  cl/.style={double=white, double distance=0.9pt, line width=0.4pt, rejgray},
  brace/.style={decorate, decoration={brace, amplitude=2.2pt}},
  flt/.pic={\fill[harmver, draw=white, line width=0.3pt] (90:0.15) -- (120:0.06) -- (150:0.15) -- (180:0.06) -- (210:0.15) -- (240:0.06) -- (270:0.15) -- (300:0.06) -- (330:0.15) -- (360:0.06) -- (390:0.15) -- (420:0.06) -- cycle;},
  rej/.pic={\draw[rejgray, line width=1.0pt, line cap=round] (-0.09,-0.09) -- (0.09,0.09) (-0.09,0.09) -- (0.09,-0.09);},
  dq/.style={circle, fill=black, inner sep=0pt, minimum size=1.5mm},
  cq/.style={circle, fill=tolblue, draw=white, line width=0.3pt, inner sep=0pt, minimum size=1.6mm},
  sq/.style={rectangle, fill=skyblue, inner sep=0pt, minimum size=1.4mm},
]
\begin{scope}[shift={(0,0)}]
  \node[letter] at (0,3.95) {(a)};
  \node[anchor=west, small, inkgray] at (0.33,3.87) {cultivation};
  \foreach \y in {2.95,2.65,2.35,2.05} \draw (0.75,\y) -- (4.10,\y);
  \draw[brace] (0.68,1.98) -- (0.68,3.02);
  \node[anchor=east, align=right, small] at (0.62,2.50) {color\\code};
  \draw[gate] (0.88,1.92) rectangle (1.30,3.08);
  \node[rotate=90, small] at (1.09,2.50) {inject $|T\rangle$};
  \draw[rounded corners=2pt, dashed, inkgray] (1.45,1.50) rectangle (3.95,3.28);
  \node[anchor=south east, small, inkgray] at (3.95,3.28) {$\times\,r$};
  \foreach \y in {2.95,2.65,2.35,2.05} \node[tgate] at (1.75,\y) {T};
  \draw[gate] (2.10,1.92) rectangle (3.15,3.08);
  \node at (2.625,2.50) {$\bar X$};
  \draw[cl] (2.625,1.92) -- (2.625,1.62);
  \node[anchor=west, small, rejgray] at (2.68,1.74) {$+1$};
  \foreach \y in {2.95,2.65,2.35,2.05} \node[tgate] at (3.55,\y) {T$^\dagger$};
  \pic at (1.98,2.35) {flt};
  \draw[code] (4.10,1.92) rectangle (4.75,3.08);
  \node[small, align=center] at (4.425,2.50) {ideal\\dec.};
  \draw (4.75,2.50) -- (5.15,2.50);
  \node[anchor=west] at (5.15,2.50) {$\tau$};
  \node[tgate, minimum width=0.42cm] (tT) at (1.75,0.75) {T};
  \node[tgate, minimum width=0.42cm, draw=rejgray, fill=black!8] (tS) at (2.75,0.75) {S};
  \draw[->, inkgray] (tT) -- (tS);
  \node[anchor=west, small, inkgray] at (3.10,0.75) {S proxy};
  \draw[inkgray, dotted, line width=0.5pt] (1.75,1.43) -- (tT);
\end{scope}
\begin{scope}[shift={(6.0,0)}]
  \node[letter] at (0,3.95) {(b)};
  \node[anchor=west, small, inkgray] at (0.33,3.87) {one code decides};
  \draw[brace] (0.92,2.30) -- (0.92,3.40);
  \node[anchor=east] at (0.86,2.85) {$|T\rangle^{\otimes t}$};
  \foreach \y in {3.30,3.04,2.66,2.40} \draw[tolblue] (1.0,\y) -- (1.30,\y);
  \foreach \y in {2.77,2.85,2.93} \fill[tolblue] (1.15,\y) circle (0.45pt);
  \draw[gate] (1.30,2.24) rectangle (3.35,3.46);
  \node[small, align=center] at (2.05,2.85) {Clifford\\skeleton};
  \pic at (2.95,3.20) {flt};
  \draw[->, harmver, dotted, line width=0.7pt] (3.11,3.20) -- (3.75,3.20);
  \draw (3.35,2.60) -- (3.75,2.60);
  \node[anchor=west] at (3.75,2.60) {$\tau$};
  \draw[->, inkgray, line width=0.6pt] (2.32,2.14) -- (1.25,1.80);
  \draw[->, inkgray, line width=0.6pt] (2.32,2.14) -- (3.25,1.30);
  \node[anchor=north] at (0.85,1.80) {$D(F)\mathrel{\ooalign{$=$\cr\hidewidth$/$\hidewidth\cr}}0$};
  \pic at (1.62,1.58) {rej};
  \draw[brace] (1.82,0.20) -- (1.82,1.12);
  \node[anchor=east] at (1.76,0.66) {$|T\rangle^{\otimes t}$};
  \foreach \y in {1.02,0.78,0.54,0.30} \draw (1.90,\y) -- (3.90,\y);
  \draw[gate, draw=harmver, fill=harmver!12] (2.05,0.17) rectangle (2.80,1.15);
  \node at (2.425,0.66) {$E(F)$};
  \draw[code] (3.00,0.17) rectangle (3.50,1.15);
  \node at (3.25,0.66) {$\figPi_C$};
  \draw[gate] (3.90,0.17) rectangle (4.35,1.15);
  \node at (4.125,0.66) {$V$};
  \draw (4.35,0.66) -- (4.60,0.66);
  \node[anchor=west] at (4.60,0.66) {$\tau$ or $0$?};
\end{scope}
\begin{scope}[shift={(12.0,0)}]
  \node[letter] at (0,3.95) {(c)};
  \node[anchor=west, small, inkgray] at (0.33,3.87) {one fault};
  \node[small] at (1.40,3.40) {GHZ fold};
  \coordinate (r) at (1.40,2.95);
  \coordinate (n1) at (0.80,2.45);
  \coordinate (n2) at (2.00,2.45);
  \draw[line width=0.7pt] (r) -- (n1); \draw[line width=0.7pt] (r) -- (n2);
  \foreach \x in {1.65,2.00,2.35} \draw[line width=0.7pt] (n2) -- (\x,1.88);
  \foreach \x in {0.45,0.80,1.15} \draw[harmver, line width=1.0pt] (n1) -- (\x,1.88);
  \node[circle, draw, fill=white, inner sep=0pt, minimum size=1.8mm] at (r) {};
  \node[circle, fill=black!45, inner sep=0pt, minimum size=1.5mm] at (n2) {};
  \pic at (n1) {flt};
  \foreach \x in {0.45,0.80,1.15} \node[dq, fill=harmver] at (\x,1.88) {};
  \foreach \x in {1.65,2.00,2.35} \node[dq] at (\x,1.88) {};
  \node[harmver] at (1.40,1.45) {$d=(\delta+1)/2$};
  \node[small] at (4.35,3.40) {verified cat};
  \foreach \x in {3.25,3.80,4.35,4.90,5.45} \draw (\x,2.60) -- (\x,1.98);
  \draw[tolblue, line width=1.0pt] (3.25,2.60) -- (5.45,2.60);
  \draw[harmver, line width=1.2pt] (4.62,2.60) -- (5.45,2.60);
  \foreach \x in {3.25,3.80} \node[cq] at (\x,2.60) {};
  \foreach \x in {4.90,5.45} \node[cq, fill=harmver] at (\x,2.60) {};
  \node[circle, draw=tolblue, fill=white, line width=0.7pt, inner sep=0pt, minimum size=2.0mm] at (4.35,2.60) {};
  \pic at (4.62,2.60) {flt};
  \foreach \x in {3.25,3.80,4.35,4.90,5.45} \node[dq] at (\x,1.98) {};
  \draw[skyblue, dashed, line width=0.6pt] (3.25,2.67) .. controls (3.70,2.94) and (5.00,2.94) .. (5.45,2.67);
  \node[sq] at (4.35,2.87) {};
  \node[tolblue] at (4.35,1.45) {$d=\delta$};
  \draw[rejgray, line width=0.3pt] (2.82,1.05) -- (2.82,3.60);
\end{scope}
\end{tikzpicture}
\caption{The main ideas of this paper. (a)~A cultivation circuit injects a T state into a color code and checks it $r$ times. A check applies a layer of T and T$^\dagger$ gates to the data qubits, measures the logical X operator, keeps the outcome $+1$ and undoes the T gates, which together measures the logical operator $TXT^\dagger$. An ideal decoder ends the circuit. Faults (stars) can occur at every operation. The S proxy replaces every T gate by an S gate, which makes the circuit Clifford but turns an X error into a Y error, whose Z part the code detects. (b)~Replacing every T gate by an injected T state leaves a noisy Clifford circuit, the skeleton, acting on $t$ ideal T states. A fault pushed to the end of the skeleton (dotted) either fires a detector, $D(F)\neq0$, and is rejected, or acts as a Pauli error $E(F)$ on the T inputs. It is harmless exactly when the code $C$ still projects the perturbed inputs onto the target $\tau$ or onto zero (Theorem~\ref{thm:main}). (c)~In a GHZ fold, one fault spreads X over a branch of data qubits (vermilion). The T gates erase the Z part of these X errors, further faults complete a logical operator (Lemma~\ref{lem:completion}), and the check keeps at most about half its distance, $d=(\delta+1)/2$ in every fold that we certified on a code of distance $\delta\leq7$. In a verified cat grown along a path from its middle (open circle), one fault spreads X only towards one end of the path. The parity of the two ends (dashed, measured through the square) catches it, and the certified checks of this kind keep the full fault distance $d=\delta$.}
\label{fig:concept}
\end{figure*}
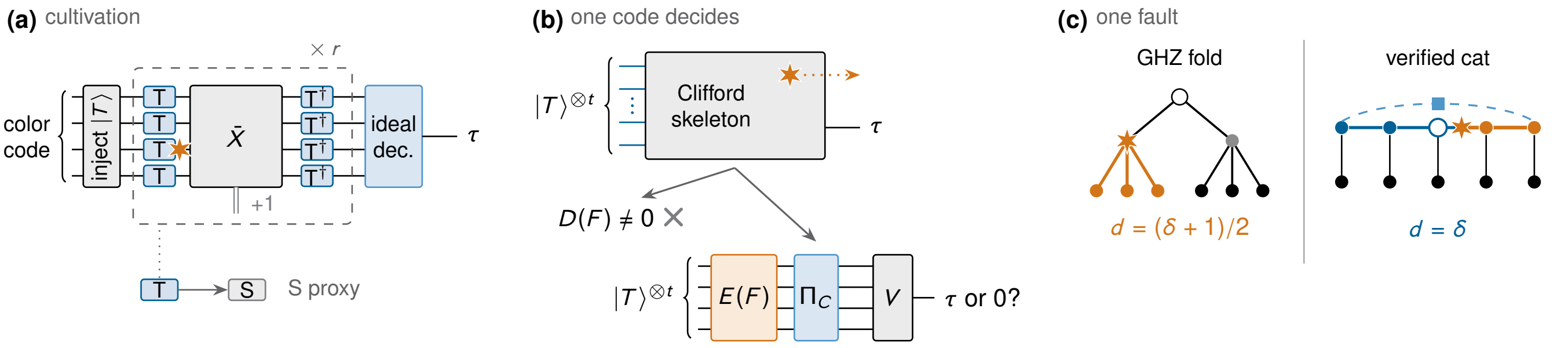

\section{Introduction}\label{sec:intro}

A fault-tolerant quantum computer needs a supply of T states, $|T\rangle=(|0\rangle+e^{i\pi/4}|1\rangle)/\sqrt2$, because the Clifford gates that error-correcting codes implement easily are not universal on their own. Magic-state cultivation~\cite{cultivation} is among the cheapest ways known to make them. It prepares a T state in a small color code and checks it repeatedly with physical T gates, keeping the state only if every check passes. Simulations predict that it reaches logical error rates that previously required distillation, at a cost comparable to that of a logical CNOT~\cite{cultivation}, and it has been demonstrated on hardware~\cite{googlecultivation,garner}. Related schemes cultivate on surface codes~\cite{chenrp2,sahay,vaknin,hartweg}, measure the check in constant depth by gauging~\cite{hetenyi}, prepare multi-qubit magic states~\cite{mincs,minmulti,hirasaki} or states beyond the third level of the Clifford hierarchy~\cite{chensqrtt}, or inject a state and verify it by postselection~\cite{li2015,chamberlandcross,chamberlandnoh,itogawa}.

Fault tolerance is understood through Pauli algebra. In a Clifford circuit a Pauli fault remains a Pauli error, and a detector error model lists which consistency checks of the circuit, its detectors, each fault triggers and which logical operators it flips~\cite{stim}. Non-Clifford gates break this picture, because a Pauli fault before such a gate becomes a non-Pauli error after it, and circuits with many of them cannot be simulated efficiently. Threshold theorems cover non-Clifford gates by building every operation from fault-tolerant gadgets~\cite{aharonovbenor,agp}, but they do not decide which faults are harmful in a given circuit. Cultivation, whose checks apply T gates directly to the data of a small code, has lacked such an exact account. Its circuits have been designed with simulations in which every T gate is replaced by a Clifford gate, and simulators that keep the T gates, which have appeared only recently~\cite{tuloup,soft,takada,merlin,magna,wanzapirain,hartweg}, evaluate one circuit at a time. Which checks are fault tolerant and why, how many T gates cultivation needs, and how to certify a check that is too large to simulate have remained open.

This paper answers these questions with an exact theory of which faults are harmful in postselected circuits with real non-Clifford gates. Its main result is that fault tolerance with T gates is still decided by Pauli algebra, on a stabilizer code that the circuit defines on its T gates and, for the checks of cultivation, on the code of the encoded state. The same algebra decides which checks keep their distance and shows how to build checks that do.

The first results hold for every circuit that prepares a magic state with Clifford operations, non-Clifford gates and postselection. One stabilizer code on the T gates decides whether a set of faults is harmful, and the faults reach this code through two linear maps that a stabilizer simulation computes (Theorem~\ref{thm:main}). The same condition decides magic-state distillation, so cultivation and distillation differ only in how faults reach the code. As consequences, an exact test decides each set of faults when the checks are deterministic, as in cultivation, at a cost polynomial in the number of T gates once the Z patterns that the code accepts are known (Theorem~\ref{thm:gauss}). Z propagation gives these patterns under mild conditions, and a transform of exponential size gives them in general. A circuit that tolerates $d-1$ faults needs at least as many T gates as a distillation protocol that suppresses errors to order $d$ needs noisy inputs (Theorem~\ref{cor:exact}). The next results concern checks on an encoded state, the setting of cultivation. Measurement gadgets that are fault tolerant in the usual sense keep the full distance, whatever single-qubit gates the check uses (Theorem~\ref{thm:strong}). A weaker condition on the code is necessary (Lemma~\ref{lem:completion}), and a check in which single faults spread X onto pairs of qubits of a logical operator keeps at most about half its distance (Corollary~\ref{cor:half}). Between the sufficient and the necessary condition, the exact fault distance is a question of Pauli algebra on the code, which needs no simulation of T gates (Theorem~\ref{thm:pauli}). These results judge the output as the simulations of cultivation do, with an ideal decoder that discards every state with a detected error, and we show what a realistic final stage changes (Proposition~\ref{prop:final}).

We then use the theory on two current problems. The first is the S proxy, the replacement of every T gate by an S gate, its Clifford square, with which cultivation has been designed and benchmarked~\cite{cultivation}. Simulations with real T gates found that the published circuits of distance three and five fail more often and tolerate fewer faults than their proxies~\cite{tuloup,chan,wanzapirain}, and Chan \emph{et al.} restored their distance with extra flag qubits and, at distance five, a further round of stabilizer measurements~\cite{chan}. The theory explains these findings and shows how general they are. On an encoded state, T and S gates differ only on X errors that pass through T gates. An S gate turns such an error into a Y error, whose Z part the code detects, while a T gate turns it into a superposition from which the code can keep the bare X error. For a single check on an ideally encoded state, the proxy can therefore only overstate the fault distance (Theorem~\ref{thm:proxybound}), by any amount, since it does not see how far a fault spreads X (Proposition~\ref{prop:blind}), and the checks in use today, in which a single fault spreads X over several data qubits, tolerate at most $(\delta-1)/2$ faults on a code of distance $\delta$, about half of what their proxies promise. We certify this on the published color-code and surface-code circuits, including the one run on hardware~\cite{googlecultivation}, and we bound it at larger distance from the structure of the check alone. The second problem is the design of checks. The theory gives checks that tolerate the full number of faults with nearest-neighbor gates on a square grid, certified up to code distance nine. Section~\ref{sec:setting} defines the setting, Secs.~\ref{sec:criterion} to~\ref{sec:structure} develop the theory, Secs.~\ref{sec:proxy} and~\ref{sec:design} apply it, and the appendices contain the proofs, algorithms and circuits.

\section{Cultivation and fault distance}\label{sec:setting}

We study circuits built from Clifford gates, which map Pauli operators to Pauli operators under conjugation, from non-Clifford gates such as the T gate, and from measurements whose outcomes are postselected, so that a run is kept only if every outcome takes its expected value. The target is a magic state $\tau$, and we assume that no Pauli operator other than the identity leaves it unchanged, as holds for $|T\rangle$ and $|\CCZ\rangle$.

Cultivation encodes the target in a stabilizer code, the joint eigenspace of a group of commuting Pauli operators called stabilizers. A logical operator is a product of Pauli operators that commutes with every stabilizer without being one, and the code distance $\delta$ is the least number of qubits on which a logical operator acts. A set of exactly $\delta$ qubits that carries a logical operator is a minimum-weight logical support. A code is of CSS type if each stabilizer is a product of X operators or a product of Z operators, and self-dual if its X and Z stabilizers have the same supports. Cultivation then checks the encoded state repeatedly. A check measures a logical operator of which the target is an eigenstate, and a state that fails a check is discarded. Each measurement is carried out by a gadget, a small circuit of ancilla qubits and gates whose measured ancillas give the outcome. A circuit is correct if, without faults, it outputs the target.

The circuits in use, and our examples, encode a T state in a triangular color code. Its $n=(3\delta^2+1)/4$ data qubits lie on a lattice of hexagonal and square faces, each face carries an X and a Z stabilizer, and X and Z on all data qubits are logical operators. A check applies a layer $U^\dagger$ of T and T$^\dagger$ gates to the data, measures the logical X operator and applies the inverse layer $U$. The pattern of T and T$^\dagger$ gates is chosen so that the check measures the logical operator $TXT^\dagger$, of which $|T\rangle$ is an eigenstate~\cite{chan}. The circuit ends in an ideal decoder, which measures all stabilizers without error, discards the state if any of them is violated, and decodes the logical qubit. It stands for the later stages of a protocol, in which the code is grown and protected by error correction, as in the simulations of Ref.~\cite{cultivation}. Section~\ref{sec:structure} shows what a noisy final stage changes, and Ref.~\cite{mehta} analyzes its cost.

Every preparation, gate, measurement and idle step of the circuit is a location, and each location fails independently with probability of order $p$. A fault is a Pauli error after a location or a wrong measurement outcome. A set of faults, a fault configuration, is \emph{harmless} if the circuit then discards the state or outputs the correct one, and \emph{harmful} if it can output a wrong state. The \emph{fault distance} $d$ is the least number of faults in a harmful configuration. An accepted state is then wrong with probability $c\,p^d+O(p^{d+1})$, and the coefficient $c$ of this leading term sums the harmful configurations of $d$ faults, each weighted by its probability, its probability of acceptance relative to the fault-free circuit and the infidelity of its output. If the data can fail just before the decoder, the fault distance cannot exceed $\delta$ (Proposition~\ref{prop:singleton}), and a circuit with $d=\delta$ keeps its \emph{full distance}. The proxy fault distance is the fault distance of the S proxy.

\section{One code decides}\label{sec:criterion}

To follow the T gates exactly we move them out of the way. A T gate equals a small Clifford circuit that consumes an ancilla prepared in $|T\rangle$~\cite{zhouleungchuang},
\begin{equation}
 \sqrt2\,\langle0|_a\,\mathrm{CX}_{q\to a}\,|\psi\rangle_q|T\rangle_a=T|\psi\rangle_q .
 \label{eq:gadget}
\end{equation}
Doing this for all $t$ T gates leaves a noisy Clifford circuit, the \emph{skeleton}, acting on $t$ ideal T states. Every fault can be pushed to its end, where it either contradicts an outcome fixed in the skeleton, a \emph{detector}, or acts as a Pauli error on the T states.

\begin{theorem}\label{thm:main}
Consider a circuit, such as a cultivation circuit ending in its ideal decoder, with $t$ T gates, without feedforward and with every measurement postselected, that prepares a target state $\tau$ when no fault occurs. Suppose that no Pauli operator other than the identity fixes $\tau$, as is the case for a T state. Then there are a stabilizer code $C$ on $t$ qubits and two linear maps, $D$ from fault configurations to detector patterns and $E$ from fault configurations to Pauli operators on $t$ qubits, such that a fault configuration $F$ is harmless if and only if $D(F)\neq0$ or
\begin{equation}
 \Pi_C\,E(F)\,|T\rangle^{\otimes t}\in\mathbb C\,\Pi_C|T\rangle^{\otimes t},
 \label{eq:collapse}
\end{equation}
where $\Pi_C$ is the projection onto $C$. The map $E$ is defined up to phases and stabilizers of $C$.
\end{theorem}

A configuration is thus harmless exactly when it is detected, or when $C$ still projects the perturbed T states onto the target or onto zero [Fig.~\ref{fig:concept}(b)]. Equation~\eqref{eq:collapse} is the condition that decides magic-state distillation, in which perfect Clifford circuits measure a code on noisy T states~\cite{distillationpaper}. Cultivation therefore differs from distillation only through $D$ and $E$, which carry faults onto the T states as a detector error model does for Clifford circuits~\cite{stim}. The theorem holds for any non-Clifford gates, each injected from its own resource state, and the fault distance of a circuit whose later operations depend on earlier outcomes (feedforward) is at most those of the circuits of its accepted runs, with equality under a mild condition (Propositions~\ref{prop:gates} and~\ref{prop:branch}, Appendices~\ref{sm:proofs} and~\ref{sec:formal}).

\paragraph*{X errors on the T states.}
What distillation does not see is the X part of $E(F)$. A Z error on a T state maps it to an orthogonal state, but an X error does not,
\begin{equation}
 X|T\rangle=\tfrac{1}{\sqrt2}\bigl(|T\rangle+iZ|T\rangle\bigr).
 \label{eq:split}
\end{equation}
An error whose X part reaches several T states is a superposition of Z patterns on them, and it is harmless if each of them is (Lemma~\ref{lem:shadow}). In particular, a code that tolerates Z errors on fewer than $d$ inputs tolerates every Pauli error on them.

\section{An exact test}\label{sec:test}

Condition~\eqref{eq:collapse} involves $|T\rangle^{\otimes t}$, but in cultivation it is cheap to decide, because the checks are deterministic. A Z error on the T states leaves the encoded state unchanged, flips it, or is rejected with certainty (Theorem~\ref{thm:dcauto}). By Eq.~\eqref{eq:split}, an error with X part on $|a|$ T states splits into $2^{|a|}$ Z patterns, each sorted by the code, and adding their amplitudes gives an exact test.

\begin{theorem}\label{thm:gauss}
If the checks of the circuit are deterministic (Definition~\ref{def:dc}), then for every fault configuration $F$ with $D(F)=0$, whose error $E(F)$ has an X part on $|a|$ inputs, there are two numbers $g_0(F)$ and $g_1(F)$, each a sum $\sum_zi^{q(z)}$ over a quadratic form $q$ with values modulo four, such that $F$ is accepted with probability $2^{-|a|}(|g_0|^2+|g_1|^2)$ relative to the fault-free circuit, and the acceptance times the infidelity of its output is $2^{-|a|}|g_1|^2$. In particular, $F$ is harmful if and only if $g_1(F)\neq0$, which is decided with $O(t^3)$ operations once the Z patterns that the code accepts are known.
\end{theorem}

The accepted Z patterns follow from the propagation of Z errors when the conditions of Theorem~\ref{thm:dcauto} hold, and otherwise from one Walsh--Hadamard transform over the $2^\kappa$ terms of the Choi state of the skeleton, with $\kappa\leq27$ for our circuits (Appendix~\ref{sm:algorithm}). Such Gauss sums are evaluated by eliminating one variable at a time, in about a millisecond per configuration at distance seven. Together with a detector error model of the skeleton from the stabilizer simulator stim~\cite{stim} and a search for faults whose detector patterns cancel, this certifies fault distances with real T gates by the same search as for Clifford circuits. A second implementation agrees fault by fault on the published distance-three circuit (Appendix~\ref{sm:algorithm}).

\section{The cost in T gates}\label{sec:tcount}

A distillation protocol has order $d$ if its output error is of order $p^d$ when each noisy input fails with probability $p$. Z faults on the T gates never fire a detector, $D=0$, although the code $C$ can reject them, and they act on $C$ like noise on the inputs of a distillation protocol, so a cultivation circuit of fault distance $d$ contains a distillation protocol of order $d$, and conversely.

\begin{theorem}\label{cor:exact}
Every cultivation circuit of fault distance $d$ for one T state, with noisy T gates, contains at least as many T gates as the least number of noisy inputs of a distillation protocol of order $d$. This number of T gates suffices if the circuit ends in an ideal decoder for its output code, and for $d\geq2$ it is at least $\max(2d,d+7)$~\cite{distillationpaper}, at least $3d-4$ for odd $d$ and at least $3d-6$ for even $d$.
\end{theorem}

Fault distance two therefore needs exactly nine T gates~\cite{distillationpaper}, attained with eleven qubits (Proposition~\ref{prop:t9}), while a single check on the color code of distance $\delta$ uses $(3\delta^2+1)/2$. The fault distance lies between $\min(d_C,d_{\mathrm{sp}})$ and $d_C$, where $d_C$ is the least number of harmful Z faults on the T gates and $d_{\mathrm{sp}}$ the least number of undetected faults whose Pauli operator on the T states is heavier than their number (Theorem~\ref{thm:attain}). What limits cultivation is how its checks spread faults, the subject of the next section.

\section{Fault tolerance of checks on an encoded state}\label{sec:structure}

Cultivation checks act on an encoded state through layers of single-qubit gates between Clifford measurement gadgets. For such checks the theory says when the non-Clifford gates are harmless and which faults they make dangerous. The first answer is the usual condition of fault tolerance, that no fault may spread to several data qubits.

\begin{theorem}\label{thm:strong}
Let a correct circuit start from an ideally encoded state of a stabilizer code of distance $\delta$, end in its ideal decoder, and act on the data, without faults, only by layers of single-qubit gates and by measurements of Pauli operators with deterministic outcomes. Suppose that every configuration of $s<\delta$ faults that the measurement gadgets accept puts Pauli errors on at most $s$ data qubits in all, where the error that a gadget leaves may be multiplied by the operator that it measures. Then no configuration of fewer than $\delta$ faults is harmful, with T gates in the layers as with S gates or any other single-qubit gates.
\end{theorem}

Single-qubit gates keep each error on its qubits, so the decoder receives a superposition of Pauli errors on fewer than $\delta$ qubits, each of which is a stabilizer or is detected (Appendix~\ref{sm:strong}). This is the usual argument of fault tolerance, and the theorem shows that it survives layers of non-Clifford gates, which turn one Pauli error into a superposition of several.

With diagonal layers the proof also shows what they add. A gate $V=\mathrm{diag}(1,e^{i\theta})$ keeps the X part of an error and adds Z only where it meets an X error,
\begin{equation}
 VXV^\dagger=(\cos\theta\,I-i\sin\theta\,Z)\,X .
 \label{eq:diag}
\end{equation}
A configuration whose errors have the X part $X^A$ and the Z part $Z^B$ in all, where $X^A$ is X on every qubit of a set $A$, reaches the decoder as a superposition of the \emph{coherent family} $X^AZ^{B+w}$, with $w$ ranging over the Z patterns on the X errors that a layer acts on, and it is harmless unless the family contains a nontrivial logical operator (Theorem~\ref{thm:family}). With S gates, $\cos\theta=0$, and only one member of the family has a nonzero amplitude, the error that the proxy tests.

The family also shows what a check must avoid, namely an X error that covers part of a minimum-weight logical support before the last layer.
\begin{lemma}[Coherent completion]\label{lem:completion}
Let $\ell$ be a minimum-weight logical X support of a CSS code of distance $\delta$, and let $u$ be a nonempty proper subset of $\ell$ that contains the support of no Z stabilizer, as holds for every such $u$ in a color code. Suppose that undetected faults leave the error $X^u$ on the data just before the last layer of diagonal gates $\mathrm{diag}(1,e^{i\theta_q})$, that further faults put X on the rest of $\ell$ immediately after this layer, and that only stabilizers are measured between this layer and the ideal decoder. Then the decoder accepts with probability $\prod_{q\in u}\cos^2\theta_q$ relative to the fault-free circuit, and it outputs the target with the logical X operator applied.
\end{lemma}

The decoder removes every member with a Z part on $u$, which is neither a stabilizer nor a logical operator, and keeps the bare $X^u$, which the further faults complete (Appendix~\ref{sm:mechanism}). If $s$ undetected faults put X on $k>s$ qubits of a minimum-weight logical support before the last layer, $\delta-k$ more faults therefore complete a logical operator (Corollary~\ref{cor:hooks}). A check with the full distance must thus be \emph{fault tolerant relative to the code}, which means that no $s$ undetected faults may leave X exactly on more than $s$, but fewer than $\delta$, qubits that lie inside a minimum-weight logical support before the last layer. This condition is much weaker than ordinary fault tolerance, because few sets of four or more qubits lie inside such supports at larger distance (Table~\ref{tab:cover}), but it is not sufficient. One fault that leaves X on a face lowers the fault distance, on the color codes at least up to $\delta=11$ (Corollary~\ref{cor:face}).

The loss is largest when single faults spread X to pairs of qubits of one logical support.

\begin{corollary}[Half the distance]\label{cor:half}
In the setting of Lemma~\ref{lem:completion}, let $\delta$ be odd, and let $(\delta-1)/2$ single faults together leave X on disjoint pairs of qubits of one minimum-weight logical support just before the last layer, without being detected, and let the remaining qubit of the support have a fault location immediately after the layer. Then the fault distance with T or T$^\dagger$ gates in the layer is at most $(\delta+1)/2$, while with S gates these faults are detected.
\end{corollary}

The pairs cover $\delta-1$ qubits of the support, so one further fault completes a logical operator (Corollary~\ref{cor:hooks} with $s=(\delta-1)/2$ and $|u|=\delta-1$), and Lemma~\ref{lem:completion} shows that S gates detect it. A check in which single faults can spread X in this way keeps at most about half its distance with real T gates, whatever its other properties, while its proxy detects these faults.

Between the sufficient and the necessary condition, the exact fault distance is decided by Pauli algebra. A complete check removes almost all the coherence of an earlier X error (Lemma~\ref{lem:filter}), so the dangerous configurations are \emph{late}. In a late configuration every X error that a layer acts on occurs inside the last check, and earlier faults leave only Z errors.

\begin{theorem}[Pauli certificates]\label{thm:pauli}
Let a circuit start from an ideally encoded state of a self-dual CSS code of distance $\delta$ with one logical qubit and the logical X operator $X^{\otimes n}$, such as a triangular color code, measure one check with layers of T and T$^\dagger$ gates that make the check logical (Lemma~\ref{lem:flat}), possibly among stabilizer measurements, and end in its ideal decoder, and let it satisfy conditions (i) to (iii) of Theorem~\ref{thm:dominance}, which say in essence that faults before the check leave, apart from Z errors, errors on no more data qubits than their number. Then its fault distance with real T gates is the smaller of $\delta$ and the least number of faults in a harmful late configuration. Whether a late configuration is harmful is decided from its X and Z patterns by Pauli algebra on the code (Theorem~\ref{thm:late}). A member of the coherent family survives only if its Z pattern meets every Z stabilizer inside the X pattern an even number of times.
\end{theorem}

Theorem~\ref{thm:pauli} combines Theorems~\ref{thm:late} and~\ref{thm:dominance}. Certifying such a check needs neither T gates nor Gauss sums, and we solve the resulting combinatorial problem as an integer program whose solutions are checked exactly (Appendix~\ref{sec:certificates}).

The same algebra compares T gates with their S stand-ins. A late $X^A$ reaches the decoder through the members whose Z patterns $w\subseteq A$ meet every Z stabilizer inside $A$ evenly. With S gates only $w=A$ remains, which is one of them because Z stabilizers have even weight, so every late configuration that is harmful with S gates is harmful with T gates (Lemma~\ref{lem:sproxy}).

\begin{theorem}[The S proxy is an upper bound]\label{thm:proxybound}
Under the hypotheses of Theorem~\ref{thm:pauli}, the fault distance with real T gates is at most the fault distance of the S proxy, which has S and S$^\dagger$ gates in place of the T and T$^\dagger$ gates.
\end{theorem}

The bound needs the encoded state. The nine-T circuit of Proposition~\ref{prop:t9}, which measures its code on unencoded T states, tolerates more faults with T gates than with S gates.

\paragraph*{The final stage.}
The ideal decoder discards every state with a detected error. A real protocol must eventually stop discarding states and correct their errors instead, and this caps its fault distance, whatever its checks.

\begin{proposition}\label{prop:final}
If a correct circuit ends in an ideal decoder that corrects errors on a code of distance $\delta'$ instead of rejecting them, and each output qubit has a fault location just before it, its fault distance is at most $\lfloor\delta'/2\rfloor+1$. Under the hypotheses of Theorem~\ref{thm:strong}, with the ideal decoder replaced by the correcting decoder of the same code, the fault distance is at least $\lfloor(\delta-1)/2\rfloor+1$, which is $(\delta+1)/2$ for odd $\delta$.
\end{proposition}

Checks of fault distance $d$ must therefore be followed by growth to a code of distance about $2d$ before the protocol stops discarding states (Appendix~\ref{sec:final}). The escape stage of Ref.~\cite{cultivation} does this. It grows the code into a large surface code and discards states that its decoder flags as uncertain. Whether a noisy growth stage keeps the fault distance of the checks is open.

\section{Why the S proxy misleads}\label{sec:proxy}

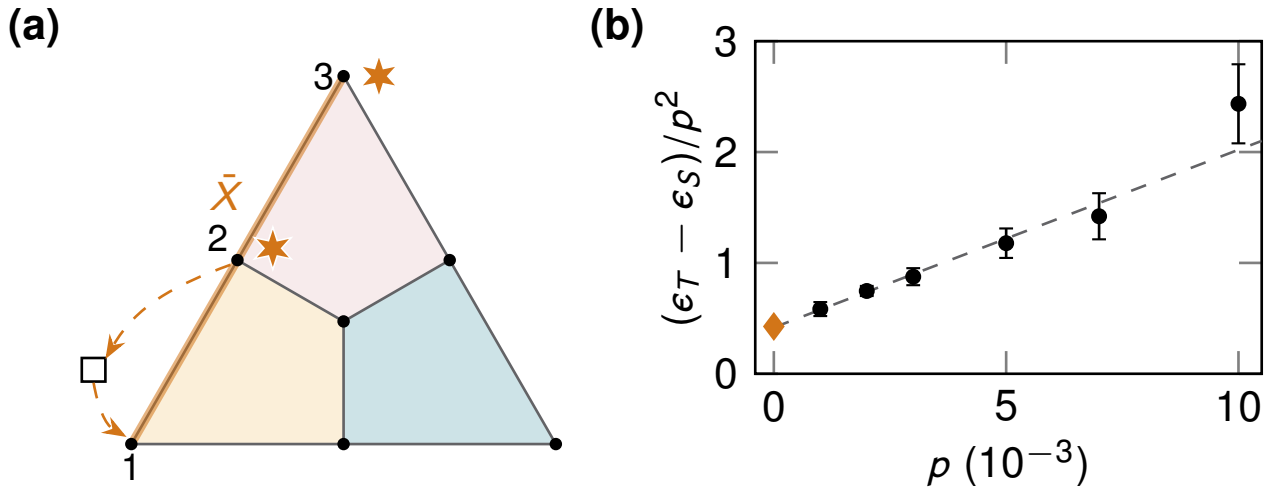
\begin{figure}[t]
\centering
\definecolor{tolblue}{cmyk}{1,0.36,0,0.30}      
\definecolor{skyblue}{cmyk}{0.63,0.23,0,0.09}   
\definecolor{harmver}{cmyk}{0,0.56,1,0.165}     
\definecolor{rejgray}{cmyk}{0,0,0,0.62}         
\definecolor{inkgray}{cmyk}{0,0,0,0.75}         
\definecolor{pauX}{cmyk}{0,0.31,1,0.10}         
\definecolor{pauY}{cmyk}{1,0,0.27,0.38}         
\definecolor{pauZ}{cmyk}{0,0.41,0.18,0.20}      

\begin{tikzpicture}[
  font=\fontsize{8}{9.6}\fontfamily{phv}\selectfont\mathversion{figsans},
  line width=0.5pt,
  >={Stealth[length=1.6mm,width=1.2mm]},
  letter/.style={font=\fontsize{8}{9.6}\fontfamily{phv}\fontseries{b}\selectfont, anchor=north west, inner sep=0pt},
  small/.style={font=\fontsize{7}{8.4}\fontfamily{phv}\selectfont\mathversion{figsans}},
]
\node[letter] at (0,3.05) {(a)};
\begin{scope}[shift={(0.55,-3.55)}, scale=0.78]
\fill[pauX!14] (0.35,4.75) -- (2.15,4.75) -- (2.15,5.79) -- (1.25,6.31) -- cycle;
\fill[pauY!14] (2.15,4.75) -- (3.95,4.75) -- (3.05,6.31) -- (2.15,5.79) -- cycle;
\fill[pauZ!14] (1.25,6.31) -- (2.15,5.79) -- (3.05,6.31) -- (2.15,7.87) -- cycle;
\draw[inkgray] (0.35,4.75) -- (3.95,4.75) -- (2.15,7.87) -- cycle;
\draw[inkgray] (2.15,4.75) -- (2.15,5.79) -- (1.25,6.31) (2.15,5.79) -- (3.05,6.31);
\draw[harmver, line width=1.6pt, line cap=round, opacity=0.55] (0.35,4.75) -- (2.15,7.87);
\fill (0.35,4.75) circle (1.5pt);
\fill (3.95,4.75) circle (1.5pt);
\fill (2.15,7.87) circle (1.5pt);
\fill (2.15,4.75) circle (1.5pt);
\fill (1.25,6.31) circle (1.5pt);
\fill (3.05,6.31) circle (1.5pt);
\fill (2.15,5.79) circle (1.5pt);
\draw[fill=white] (-0.07,5.28) rectangle (0.13,5.48);
\draw[->, harmver, dashed, line width=0.5pt] (1.19,6.27) to[out=200, in=60] (0.13,5.47);
\draw[->, harmver, dashed, line width=0.5pt] (0.03,5.27) to[out=-80, in=150] (0.31,4.81);
\node[anchor=north, inner sep=1.5pt, small] at (0.35,4.75) {1};
\node[anchor=south east, inner sep=1.5pt, small] at (1.25,6.31) {2};
\node[anchor=east, inner sep=1.5pt, small] at (2.15,7.87) {3};
\node[star, star points=6, star point ratio=2.4, fill=harmver, draw=white, line width=0.3pt, inner sep=0pt, minimum size=2.8mm] at (1.55,6.41) {};
\node[star, star points=6, star point ratio=2.4, fill=harmver, draw=white, line width=0.3pt, inner sep=0pt, minimum size=2.8mm] at (2.45,7.87) {};
\node[harmver, anchor=south east, inner sep=1pt] at (1.37,6.68) {$\bar X$};
\end{scope}
\node[letter] at (3.85,3.05) {(b)};
\begin{axis}[at={(4.95cm,0.62cm)}, anchor=south west, width=3.35cm, height=2.2cm, scale only axis,
  xmin=-0.4, xmax=10.5, ymin=0, ymax=3.0, xtick={0,5,10}, ytick={0,1,2,3},
  tick label style={font=\fontsize{8}{9.6}\fontfamily{phv}\selectfont\mathversion{figsans}}, label style={font=\fontsize{8}{9.6}\fontfamily{phv}\selectfont\mathversion{figsans}},
  xlabel={$p$ ($10^{-3}$)}, ylabel={$(\epsilon_T-\epsilon_S)/p^2$},
  axis line style={line width=0.5pt}, tick style={line width=0.5pt}, xlabel shift=-3pt, ylabel shift=-4pt,
  clip=false]
\addplot[inkgray, dashed, line width=0.5pt, domain=0:10.5] {0.4175 + 0.160295*x};
\addplot[only marks, mark=*, mark size=1.3pt, black, error bars/.cd, y dir=both, y explicit, error bar style={line width=0.5pt}] coordinates {(1.000,0.5836) +- (0,0.0631) (2.000,0.7468) +- (0,0.0455) (3.000,0.8757) +- (0,0.0768) (5.000,1.1778) +- (0,0.1329) (7.000,1.4206) +- (0,0.2079) (10.000,2.4351) +- (0,0.3566)};
\addplot[only marks, mark=diamond*, mark size=2.4pt, harmver] coordinates {(0,0.4267)};
\end{axis}
\end{tikzpicture}
\caption{How real T gates lose distance in the distance-three circuit of Ref.~\cite{cultivation}. (a)~One of its $48$ harmful pairs of faults without idle noise (stars) on the seven data qubits of the color code (faces shaded), following Ref.~\cite{chan}. The first fault, an X error on qubit~2, is copied to qubit~1 through a qubit of the fold (square). In the last layer, S gates turn $X_1X_2$ into $Y_1Y_2$, which the top face detects. T gates turn it into $\tfrac12(X+Y)_1(X+Y)_2$. The second fault, on qubit~3, completes the component $X_1X_2$ to the logical operator $\bar X$ along the left edge, which passes every check. (b)~Difference of the logical errors per accepted run with T and with S gates, divided by $p^2$, in the state-vector simulations published with the circuit~\cite{gsjcode}, under uniform noise with idling (dots, one standard deviation), with a weighted linear fit to all points as a consistency check (dashed). The diamond is the exact limit $32/75$ at $p\to0$ from Theorem~\ref{thm:main}.}
\label{fig:gsj}
\end{figure}

Theorem~\ref{thm:main} locates where T gates and the S proxy differ. On an encoded state, an error that never meets a T gate, or meets it with only a Z part, is treated alike by both, and the two differ only on X errors that pass through T gates, which split as in Eq.~\eqref{eq:split}, while the proxy keeps one definite Pauli error. Lemma~\ref{lem:completion} shows why this matters. With T gates the bare $X^u$ passes the decoder with probability $2^{-|u|}$, and the output has fidelity $1/2$ with $|T\rangle$. S gates turn it into a Y error, whose Z part the code detects even after later faults complete its X part. S and S$^\dagger$ are the only diagonal gates that remove the bare X error, so the proxy hides exactly this loss [Fig.~\ref{fig:gsj}(a)].

The checks in use today are not fault tolerant relative to their code. They measure the logical X operator with a GHZ fold, a tree of CNOTs that collects the parity of the data onto one measured qubit and is then undone. One X fault on the tree spreads, when it is undone, to a branch of data qubits, a hook [Fig.~\ref{fig:concept}(c)]. The proxy detects the hook through its Z part, but with T gates the hook can lose this Z part, and further faults complete it to a logical error. One hook and one data fault suffice at $\delta=3$ (Fig.~\ref{fig:gsj}). In general, $(\delta-1)/2$ hooks that each reach two qubits of a minimum-weight logical support and one data fault suffice, as in Corollary~\ref{cor:half}.

\begin{proposition}[computer-assisted]\label{prop:gsj}\label{prop:fold}
With real T gates in place of the S gates of their simulations, in the pattern of T and T$^\dagger$ gates that makes every check logical and, at distance five, with the Pauli frames of the growth undone before the T layers (Appendix~\ref{sm:gsj}), (i) the distance-three and distance-five cultivation circuits of Ref.~\cite{cultivation} have fault distance two and three, while their proxies have three and four, and (ii) GHZ-fold checks of the same kind on color codes of distance three, five and seven have fault distance two, three and four, for every fold tree we tried.
\end{proposition}

\begin{figure}[t]
\includegraphics[width=\columnwidth]{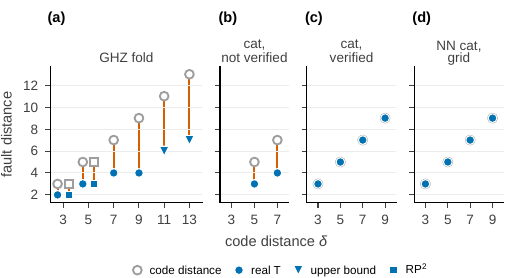}
\caption{Fault distance with real T gates (filled) and code distance $\delta$ (open), the most that the S proxy can reach. The proxy reaches $\delta$ in every case where we certified it, except the distance-five circuits of Refs.~\cite{cultivation,chenrp2}, whose proxies have fault distance four. The vermilion gaps are the distance lost to real T gates. (a)~GHZ folds, including the circuits of Ref.~\cite{cultivation}, and the surface-code scheme of Ref.~\cite{chenrp2} (squares). At $\delta=11$ and $13$ the triangles are the lowest proven upper bounds over the binary folds we evaluated, from Lemma~\ref{lem:completion} and the fold tree alone, and at $\delta=9$ the lowest of these bounds is certified for a fold without stabilizer rounds (Appendix~\ref{sec:folds}). (b), (c)~Cat-state checks without and with verification. At $\delta=9$ the verified cat in (c) is the check without stabilizer rounds, and at $\delta=7$ it keeps the full distance with and without flagged stabilizer rounds around it. (d)~Nearest-neighbor cat checks on a square grid. At $\delta=9$ both (c) and (d) are certified by the integer program of Appendix~\ref{sec:certificates}.}
\label{fig:fd}
\end{figure}

Up to $\delta=7$, every certified case has fault distance $(\delta+1)/2$ (Fig.~\ref{fig:fd}). At $\delta=3$ no fold tree does better, because every pair of data qubits lies in a minimum-weight logical support (Proposition~\ref{prop:fold3}). At $\delta=9$, $11$ and $13$, beyond the reach of simulation, Lemma~\ref{lem:completion} gives upper bounds from the fold tree alone, and the lowest of them over the binary folds we evaluated are four, six and seven, at or below $(\delta+1)/2$ (Appendix~\ref{sec:folds}). Without stabilizer rounds a fold is a single check, which Theorem~\ref{thm:pauli} decides, and the integer program of Appendix~\ref{sec:certificates} certifies two binary folds at $\delta=9$ with fault distance five and four, equal to their bounds, so half the distance is not a floor. The bounds do not decide every tree, but they show that the loss continues at larger distance. It comes from the fold, not from particular circuits or codes. The surface-code scheme of Ref.~\cite{chenrp2} prepares its state in an RP$^2$ code, a surface code on the real projective plane, and checks it on a $[[15,1,3]]$ code by contracting $X^{\otimes15}$ onto one qubit with a tree of CNOTs. Its circuits of distance three and five have fault distance two and three, against three and four for their proxies. Each of the $12$ harmful pairs at distance three is a hook of the tree plus one data fault, while at distance five the harmful triples combine a Z fault during the growth with two faults on one data qubit of the last check (Appendix~\ref{sm:gsj}).

The loss is costly, because the folds that we certified need a code of distance at least $2d-1$ for fault distance $d$, which asymptotically quadruples the qubits on which its T gates act, 61 data qubits instead of 19 for $d=5$. Its harmful configurations are suppressed only by their acceptance $2^{-|u|}$. The distance-three circuit of Ref.~\cite{cultivation} has the leading term $32p^2/75$, as in the exact series of Ref.~\cite{wanzapirain}, and the circuit run on hardware~\cite{googlecultivation} also has fault distance two, against three for its proxy. At $p=10^{-3}$, under SI1000 noise, a standard model of superconducting hardware (Sec.~\ref{sec:design}), real T gates fail about three times as often as the proxy on the distance-three circuit of Ref.~\cite{cultivation}, $5.8(1.7)$ times on its distance-five circuit, whose ratio under uniform noise is $9.4(1.2)$, and $6.1$ times on the circuit run on hardware. Projections of the hardware experiment to lower noise along a $p^3$ slope, the slope of its proxy, are therefore optimistic by a factor that grows as $1/p$ (Appendix~\ref{sm:gsj}).

The proxy cannot see how far a fault spreads X, so the loss it hides has no limit.

\begin{proposition}[The proxy is blind to X spreading]\label{prop:blind}
Under the hypotheses of Theorem~\ref{thm:pauli}, if no fault leaves a Z error on more than one data qubit and no fault after the last layer acts on more than one data qubit, the S proxy has fault distance~$\delta$.
\end{proposition}

Spreading X enters a late configuration only through the pattern that crosses the last layer, and this pattern cancels from the condition under which the proxy outputs a wrong state (Appendix~\ref{sec:proxybound}). Checks with a cat state and without stabilizer rounds satisfy the hypothesis, verified or not, because the cat qubits act on the data only as controls of CNOTs, so a Z error on the data never spreads. Their proxies keep the full distance however far one fault spreads X, as Table~\ref{tab:full} shows for verified and unverified cats alike, also with stabilizer rounds around them. With Lemma~\ref{lem:completion}, the proxy can therefore overstate the fault distance by any amount. Grow an unverified cat along a path from one end, couple its last $\delta-1$ sites to $\delta-1$ qubits of a minimum-weight logical support, and read it out without undoing the growth. One fault on the first of these sites, before it grows the next, spreads X onto all of them, and one data fault completes the logical operator. The real-T fault distance is then at most two at every distance $\delta$, while the proxy has $\delta$.

By Theorem~\ref{thm:proxybound}, such losses are the only way in which the proxy errs for a single check on an ideally encoded state. The published circuits, with injection, growth and several checks, lie outside the hypotheses of the theorem, but every encoded circuit that we certified obeys its bound. Without encoding, the proxy can also understate fault tolerance. A code measured directly on T states can reject a Z error through the interference of its T inputs, which S inputs lack, so the nine-T circuit of Proposition~\ref{prop:t9} has fault distance two, but one with S or Z gates. A variant of it answers in the negative the question of Ref.~\cite{simsek}, whether a circuit whose S and Z versions are not fault tolerant must also fail to be fault tolerant with T gates (Appendix~\ref{sm:t9}).

\section{Checks that keep the full distance}\label{sec:design}

By Theorem~\ref{thm:strong}, the full distance is kept if no fault spreads X over several data qubits. A Shor-style check~\cite{shor} prepares a cat state $(|0\cdots0\rangle+|1\cdots1\rangle)/\sqrt2$ on fresh qubits, verifies it by measuring the parities of pairs of its qubits, couples each cat qubit to one data qubit by a CNOT and reads the outcome of the check from the parity of the cat qubits in the X basis. With the cat verified $\delta-1$ times it meets the hypothesis of the theorem at every code distance, and the check itself fits on a square grid with nearest-neighbor gates (Corollary~\ref{cor:shor}). The verification is needed, since at $\delta=5$ every check with an unverified cat grown along a path, from one end or from a root with at least two sites on each side, and read out without undoing the growth loses distance (Proposition~\ref{prop:pairs}).

\begin{figure}[t]
\includegraphics[width=\columnwidth]{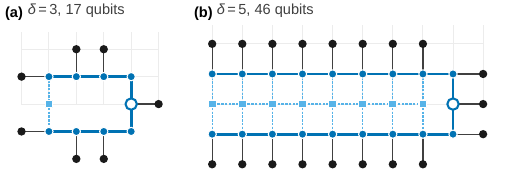}
\caption{Nearest-neighbor cat checks on a square grid, with data qubits (black), the cat path (blue) grown from its middle (open circle), and verification ancillas (light blue) with the parities they measure (dashed). (a)~The ring at $\delta=3$, whose single ancilla measures the parity of the two ends. (b)~The ladder at $\delta=5$, whose eight ancillas between the arms measure the rungs and diagonals. The three data qubits to the right sit next to the turn of the path, the middle one next to the root.}
\label{fig:nn}
\end{figure}

Verifying $\delta-1$ times is costly, and Theorem~\ref{thm:pauli} shows that one verification suffices at the distances that we can certify. On a square grid we grow the cat along a path from its middle, so that a fault on a path qubit spreads X only towards the end beyond it, and the parity of the two ends catches every such spread [Fig.~\ref{fig:concept}(c)]. Each data qubit sits next to its own path qubit, and after the coupling the growth is undone, so that the other path qubits return to $|0\rangle$ and give further detectors. At $\delta=3$ the path is a ring. At $\delta=5$ it doubles back to form a ladder on three rows, and ancillas in the middle row measure the parities across its two arms, the rungs, and along their diagonals (Fig.~\ref{fig:nn}).

\begin{proposition}[computer-assisted]\label{prop:vcat}\label{prop:nn}
With real T gates and all-to-all connectivity, checks with a verified cat keep the full fault distance three, five and seven on the color codes of distance three, five and seven, and nine on the distance-nine code without stabilizer measurements around the check ($182$ qubits). Without verification, the checks at distance five and seven have three and four. With nearest-neighbor gates on a square grid, the ring check on the distance-three code ($17$ qubits) and the ladder check on the distance-five code ($46$ qubits) have fault distance three and five, and the ladders on the distance-seven and distance-nine codes ($91$ and $151$ qubits) have seven and nine, equal to the code distance in each case.
\end{proposition}

The certificates at $\delta=9$ come from Theorem~\ref{thm:pauli} alone, with an integer program whose infeasibility below nine faults is established by a floating-point solver rather than by exhaustive search. A second solver confirms the ladder at $\delta=9$ and the verified cat at $\delta=7$, and at $\delta=7$ exact searches that use neither the integer program nor the theorem confirm both checks.

The full distance pays off when the waiting data collect little noise. We use SI1000 noise~\cite{si1000}, in which two-qubit gates fail with probability $p$, measurements with $5p$, resets with $2p$, single-qubit gates and idle qubits with $p/10$, and qubits idle during a measurement or reset with $2p$, and we gather resets and measurements into a few shared layers. From the same ideally encoded state, at $p=10^{-3}$, the ring accepts 89\% of runs and fails with probability $1.6\times10^{-8}$ per accepted run, fourteen times less often than the best GHZ fold that we built and fifty times less often than the published distance-three circuit, which also prepares its state. At $\delta=5$ the ladder fails with $8.6(1.0)\times10^{-11}$ and a verified cat with all-to-all gates with $3.7(1.2)\times10^{-12}$, against $1.63(5)\times10^{-10}$ for the unverified cat and $2.52(5)\times10^{-10}$ for the fold, with one standard error of the estimator in parentheses (Appendix~\ref{sm:nn}). These numbers compare checks on the same ideal input, not complete factories. Our complete nearest-neighbor protocol at $\delta=3$, which prepares its own state, fails with $4.7\times10^{-6}$, more often than the published distance-three circuit ($8.0\times10^{-7}$), because of the stabilizer measurements that it needs (Appendix~\ref{sec:complete}).

\section{Discussion}\label{sec:discussion}

Fault tolerance with real T gates is still decided by Pauli algebra. For any circuit that prepares a magic state, one stabilizer code on the T gates decides it, as it decides distillation, and for the checks of cultivation the code of the encoded state decides it, without any simulation of T gates. Theorem~\ref{thm:main} holds for any non-Clifford gates (Proposition~\ref{prop:gates}) and Theorem~\ref{cor:exact} for any target, so the theory also covers the cultivation of other magic states~\cite{mincs,minmulti,hirasaki,chensqrtt}. Applied to the circuits in use, the same algebra shows that they tolerate at most half as many faults as their code allows, on color and on surface codes, while their S proxies promise more, and it yields checks that keep the full distance. The theory gives five rules for designing cultivation.
\begin{enumerate}\setlength{\itemsep}{0pt}\setlength{\parskip}{0pt}
\item Certify fault tolerance with real T gates. For a single check on an encoded state the S proxy can only overstate it, by any amount, and without encoding it can also understate it.
\item Before the last T layer, let no $s$ undetected faults leave X on more than $s$ qubits inside a minimum-weight logical support, and no single fault leave X on a face. The GHZ folds and unverified cats that we certified violate this.
\item Verify the cat state. With nearest-neighbor gates, a cat grown along a path from its middle and verified by the parity of its ends, and at $\delta\geq5$ also by rungs and diagonals, keeps the full distance in the ring and ladder checks that we certified, up to $\delta=9$.
\item Gather the resets and measurements of a check into a few shared layers, so that the waiting data collect less noise.
\item Keep discarding states with detected errors until the code has grown to about twice the target fault distance, as the escape stage of Ref.~\cite{cultivation} does.
\end{enumerate}
Our certifier applies to any cultivation circuit written for stim with S gates standing for T gates, the convention of Ref.~\cite{cultivation}. The loss is not specific to color codes. The RP$^2$ surface-code scheme loses distance too, through hooks of its contraction tree at distance three and through a fault of its growth stage at distance five, while the fold-transversal surface-code scheme of Ref.~\cite{sahay} need not, as simulations suggest~\cite{hartweg}, and the criterion can decide both. Our results assume stochastic Pauli noise. Coherent errors and leakage~\cite{googlecultivation,hines} are not covered by our fault distances, although coherent errors enter Theorem~\ref{thm:main} as superpositions of Pauli operators (Appendix~\ref{sec:othergates}). A complete nearest-neighbor protocol at distance three keeps fault distance three, but its stabilizer measurements dominate its cost (Appendix~\ref{sec:complete}). A noisy growth stage, cheaper stabilizer measurements on the grid, shorter full-distance checks and lower bounds on their gates are the main open problems.

\begin{acknowledgments}
This work was carried out with the assistance of the language model Claude Opus 5.5. The final content, analysis and conclusions remain the sole responsibility of the authors. Computations used stim~\cite{stim}, Gurobi and HiGHS. The access to the computational infrastructure of the OP VVV funded project CZ.02.1.01/0.0/0.0/16\_019/0000765 ``Research Center for Informatics'' is also gratefully acknowledged. This work has been supported by the European Union's HORIZON-MSCA-2023-DN-JD programme under the Horizon Europe (HORIZON) Marie Sk\l{}odowska-Curie Actions, grant agreement 101120296 (TENORS).
\end{acknowledgments}

\paragraph*{Data availability.}
The code, circuits and numerical results that support this study are available from the authors upon reasonable request.

\appendix

\section{Protocols, faults and decoders}\label{sec:formal}

\paragraph*{Circuits and protocols.}
Throughout, $|T\rangle=(|0\rangle+\omega|1\rangle)/\sqrt2$ with $\omega=e^{i\pi/4}$. A \emph{Clifford+T protocol} acts on qubits initialized in $|0\rangle$, uses Clifford gates, T and T$^\dagger$ gates (both counted as T gates), Pauli measurements, classical randomness, feedforward, postselection and discarding, and outputs $r$ qubits. Its target is a pure state $\tau$ with stabilizer nullity $\nu(\tau)=r$, meaning that no Pauli operator other than the identity fixes $\tau$, as holds for $|T\rangle$ and $|\CCZ\rangle$. A protocol is \emph{non-adaptive and fully postselected} if it is a fixed sequence of Clifford gates, T and T$^\dagger$ gates, projections $(I+sQ)/2$ onto Pauli operators $Q$ with fixed signs $s$, and removals $\langle k|_q$ with fixed $k\in\{0,1\}$. Such a circuit applies a single Kraus operator $K$ and is \emph{correct} if $K|0\rangle$ is a nonzero multiple of $|\tau\rangle$.

\paragraph*{Faults.}
Every preparation, gate and measurement not declared ideal is a \emph{location}, while removals are ideal. A fault at a location is either a Pauli operator on its support (immediately after a preparation or gate, or immediately before a projection) or a flip of a projection's sign. The \emph{fault space} $\fs$ is the $\F$-vector space of assignments of Pauli operators and flips to locations, and the weight $\wt(F)$ of $F\in\fs$ is the least number of single faults that compose to $F$ (counting a Pauli fault and a flip at the same measurement twice). Each location suffers each of its faults independently with probability $p\lambda>0$, with constant $\lambda$ per fault type. A run of an adaptive protocol is specified by its random choices and outcomes $b$ together with the faults $F$ at the executed locations, giving a Kraus operator $K_{b,F}$. The pair $(b,F)$ is \emph{harmless} if $K_{b,F}|0\rangle\in\mathbb C|\tau\rangle$, which includes rejection. The postselected output error is $\epsilon=O(p^d)$ if and only if every run with $\wt(F)<d$ is harmless and some fault-free run is accepted. The largest such $d$ is the \emph{fault distance}, and harmful configurations of weight $d$ are \emph{malignant}.

\paragraph*{Ideal decoders.}
Encoded outputs are judged after an ideal decoder consisting of projections onto the stabilizers of an $[[m,r,\delta]]$ code $Q$ on $m$ output qubits followed by an ideal Clifford unencoding, as in simulations of cultivation~\cite{cultivation}. The decoder is \emph{admissible} if the fault-free state at its input lies in $Q$, so that it accepts the fault-free run with certainty. Without admissibility the decoder could perform T-dependent checks itself, and nine noisy T states handed to an ideal decoder of the code of Proposition~\ref{prop:t9} would already tolerate every single fault.

\begin{proposition}\label{prop:singleton}
If a correct circuit ends with an admissible ideal decoder for an $[[m,r,\delta]]$ code, and each of the $m$ qubits has an idle location immediately before the decoder, then its fault distance $d$ is at most $\delta$, and $m\geq r+2(d-1)$.
\end{proposition}

\begin{proof}
A logical Pauli of weight $\delta$ placed at the idle locations commutes with the stabilizers, passes the decoder and applies a nontrivial Pauli to the decoded state, which differs from $|\tau\rangle$ because $\nu(\tau)=r$. The inequality is the quantum Singleton bound~\cite{knilllaflamme}.
\end{proof}

\begin{proposition}\label{prop:branch}
Let a Clifford+T protocol have fault distance $d$, and let $b$ be any accepted fault-free run. The non-adaptive, fully postselected circuit formed by the operations executed along $b$, with every measurement replaced by the projection onto its outcome in $b$ and every discarded qubit removed with an outcome of nonzero amplitude, is correct and has fault distance at least $d$.
\end{proposition}

\begin{proof}[Proof of Proposition~\ref{prop:branch}]
For stochastic Pauli noise, the unnormalized accepted output is $\sum_{b',F'}q_{b',F'}\,K_{b',F'}\rho_0K_{b',F'}^\dagger$, summed over accepted runs $b'$ and fault configurations $F'$ along them with probabilities $q_{b',F'}>0$. Since a sum of positive operators is a multiple of $|\tau\rangle\langle\tau|$ if and only if each term is, a run $(b,F)$ is harmless exactly when its own term is. Fixing an accepted fault-free run $b$ fixes all random choices and outcomes, so $K_{b,F}$ is the Kraus operator of a non-adaptive circuit with $F$ inserted. A discarded qubit contributes $\sum_k\langle k|\cdot|k\rangle$, so replacing the partial trace by $\langle k|$ for an outcome $k$ with $\langle k|K_{b,\emptyset}|0\rangle\neq0$ keeps the circuit correct and every harmless configuration harmless. Hence every configuration with $\wt(F)<d$ is harmless in the circuit.
\end{proof}

Thus every T-count lower bound for correct non-adaptive circuits of fault distance $d$ holds for every accepted fault-free run of an adaptive protocol of fault distance $d$. Conversely, if every record accepted with faults is also accepted without faults with nonzero probability, the protocol's fault distance equals the least fault distance of the circuits of its accepted fault-free runs.

\section{The criterion}\label{sm:proofs}

\begin{proof}[Proof of Theorem~\ref{thm:main}]
The proof has five steps. We build the skeleton, show that the signs of its projections act as Pauli operators, move the faults onto the T inputs, derive the collapse condition, and finally treat Z faults at T gates.

\emph{Skeleton.} Equation~\eqref{eq:gadget} follows from the state $\mathrm{CX}_{q\to a}(\alpha|0\rangle+\beta|1\rangle)(|0\rangle+\omega|1\rangle)/\sqrt2$, whose component with $a$ in $|0\rangle$ is $(\alpha|0\rangle+\beta\omega|1\rangle)/\sqrt2$. We introduce one ancilla $a_j$ per T gate, prepared ideally in $|T\rangle$, and write T$^\dagger=S^\dagger T$ with an ideal $S^\dagger$. A fault of the $j$th T gate acts after the gadget. The circuit becomes $K_F=2^{t/2}K_{S,F}(|0\rangle\otimes|T\rangle^{\otimes t})$. Here the skeleton $K_{S,F}$ consists of Clifford gates, projections and removals, and $F$ is inserted at the same qubits and times. Let $M_F=K_{S,F}(|0\rangle\otimes\cdot)$, a linear map from $t$ qubits to $r$ qubits, and let $M=M_0$. As in the proof of Theorem~1 of Ref.~\cite{distillationpaper}, $M$ applied to half of $t$ Bell pairs gives a stabilizer state on $t+r$ qubits. By the local-Clifford normal form of bipartite stabilizer states, $M=c\,V\Pi_C$ with a stabilizer code $C$ on $t$ qubits and a Clifford isometry $V$ from $C$ onto its range. The range contains $|\tau\rangle$ and is a stabilizer subspace. Since no Pauli operator other than the identity fixes $\tau$, the range is the whole output space. So $\dim C=2^r$, and $V$ is unitary from $C$ onto $(\mathbb C^2)^{\otimes r}$.

\emph{Signs.} We treat each removal $\langle k|_q$ as a projection of $Z_q$ with sign $(-1)^k$ followed by an ideal removal. Let $\sigma\in\F^\mu$ list the signs of all $\mu$ projections relative to the circuit, so that $M^{(0)}=M$. Let $|\Psi^{(\sigma)}\rangle=(M^{(\sigma)}\otimes I)\sum_x|x\rangle|x\rangle$ be the unnormalized Choi vector. If we follow the circuit with the stabilizer formalism~\cite{aaronsongottesman}, the Choi vector stays a pure stabilizer state. The Pauli of every projection therefore either anticommutes with a current stabilizer or is one up to sign. In the first case the projection halves the squared norm and replaces that stabilizer by the projected Pauli with the chosen sign. In the second case it keeps the state or annihilates it, depending on the sign. The unsigned stabilizer groups and the norms of the nonzero $|\Psi^{(\sigma)}\rangle$ are therefore independent of $\sigma$. The signs required by the deterministic projections are affine functions of the earlier signs, so $|\Psi^{(\sigma)}\rangle\neq0$ exactly when $D_0\sigma=0$ for a linear map $D_0$. Two stabilizer states with the same unsigned group and norm differ by a Pauli operator and a phase. This Pauli operator is determined by the sign difference modulo the stabilizer group, and it depends on it linearly. Hence, for $D_0\sigma=0$, $|\Psi^{(\sigma)}\rangle=\zeta_\sigma(A_\sigma\otimes B_\sigma)|\Psi^{(0)}\rangle$ with $|\zeta_\sigma|=1$ and Pauli operators $A_\sigma$ on the output and $B_\sigma$ on the reference. That is, $M^{(\sigma)}=\zeta_\sigma A_\sigma MB_\sigma^{\mathsf T}$.

\emph{Faults.} We move a Pauli $P$ inserted after an operation to the end of the circuit. A Clifford gate $G$ turns it into $GPG^\dagger$. A projection $(I+sQ)$ with which it anticommutes becomes $(I-sQ)$ as $P$ passes it. A removal absorbs the Z part of $P$ on the removed qubit, and its projection turns the X part into a sign flip. A sign flip changes $\sigma$ directly. Hence $M_F=\zeta\,P_{\mathrm{out}}(F)\,M^{(s(F))}$ with a phase $\zeta$, a Pauli $P_{\mathrm{out}}(F)$ on the output and a sign vector $s(F)$, both linear in $F$. We set $D(F)=D_0s(F)$, where $D_0$ has one row per projection, nonzero only for the projections whose outcome is fixed in the skeleton. If $D(F)\neq0$, then $M_F=0$ and the configuration is rejected. Otherwise $M_F=\zeta'P_{\mathrm{out}}A_{s}MB_{s}^{\mathsf T}$. It remains to move the output Pauli onto the inputs. For an output Pauli $R$, write $V\Pi_C=W(I\otimes\langle0|^{\otimes(t-r)})U\Pi_C$, with a Clifford unitary $U$ on $t$ qubits that maps $C$ onto $(\mathbb C^2)^{\otimes r}\otimes|0\rangle^{\otimes(t-r)}$ and a Clifford unitary $W$ on $r$ qubits. Then $RM=ML$ with the Pauli $L=U^\dagger(W^\dagger RW\otimes I)U$, which commutes with the stabilizers of $C$. Hence $M_F=\zeta''M\,E(F)$ with $E(F)=L(P_{\mathrm{out}}A_s)B_s^{\mathsf T}$. Finally, $ME\propto ME'$ holds exactly when $E'E^\dagger$ acts on $C$ as a scalar, that is, when it lies in $\langle iI,S_C\rangle$. This equivalence holds because $V$ is injective on $C$, and because a Pauli that anticommutes with a stabilizer maps $C$ to an orthogonal code. So $E$ is well defined and linear modulo $\langle iI,S_C\rangle$ on the configurations with $D(F)=0$, and we extend it linearly to all of $\fs$.

\emph{Collapse.} With these maps, $K_F|0\rangle=2^{t/2}\zeta''c\,V\Pi_CE(F)|T\rangle^{\otimes t}$ and $K_0|0\rangle\propto V\Pi_C|T\rangle^{\otimes t}\propto|\tau\rangle$. Since $V$ is injective on $C$, $K_F|0\rangle\in\mathbb C|\tau\rangle$ exactly when Eq.~\eqref{eq:collapse} holds.

\emph{Z faults at T gates.} A Z fault after the gadget of the $j$th T gate on qubit $q$ satisfies $Z_q\langle0|_a\mathrm{CX}_{q\to a}=\langle0|_aZ_aZ_q\mathrm{CX}_{q\to a}=\langle0|_a\mathrm{CX}_{q\to a}Z_a$. Here we used $\langle0|_a=\langle0|_aZ_a$ and $\mathrm{CX}_{q\to a}Z_a=Z_qZ_a\mathrm{CX}_{q\to a}$. The fault thus acts as $Z$ on the ideal input $a_j$, so $D=0$ and $E=Z_j$. A Z fault before the gate gives the same result, because Z commutes with T and with $S^\dagger$.
\end{proof}

The proof shows in addition that $\dim C=2^r$ with $\nu(\tau)=r$, $\Pi_C|T\rangle^{\otimes t}\neq0$, $M=c\,V\Pi_C$ with $V\Pi_C|T\rangle^{\otimes t}\propto|\tau\rangle$, and a Z fault immediately before or after the $j$th T gate has $D=0$ and $E=Z_j$. The pair $(D,E)$ plays the role of a Clifford detector error model~\cite{stim,delfossepaetznick,bombinzx}, with logical observables replaced by Pauli operators on the T inputs and tested by Eq.~\eqref{eq:collapse} (Appendix~\ref{sm:algorithm}). Note that a measurement whose outcome is deterministic in the circuit need not be a detector of the skeleton. When its outcome depends on the T inputs, as for checks after a T layer, its projection enters $C$ rather than $D$ and rejects faults only through Eq.~\eqref{eq:collapse}. In the $\delta=3$ circuit of Ref.~\cite{cultivation}, 4 of the 20 projections are of this kind, matching the 16 Pauli-web detectors found in Ref.~\cite{wanzapirain}, whereas all 20 are detectors with S gates.

\paragraph*{Related descriptions.}
Replacing T gates by gadgets is standard in simulation~\cite{bravyigosset}. Earlier cultivation analyses propagate faults forward through the non-Clifford gates as Clifford errors~\cite{surti,chan}, diagonal Clifford operators~\cite{simsek}, or stabilizer and tensor-network decompositions~\cite{soft,clifft,symft,wanzhong,takada,wanzapirain,syqma}. Theorem~\ref{thm:main} instead pulls faults back to the T inputs, where a single fault-independent stabilizer code $C$ judges a Pauli operator $E(F)$ linear in $F$.

\subsection{X errors on the inputs}

For a Pauli operator $E$ on the T inputs, let $|E|_S$ be the least number of qubits on which an element of $E\langle iI,S_C\rangle$ acts.

\begin{lemma}\label{lem:shadow}
(i)~If every pattern $v+w$ with $\operatorname{supp}w\subseteq\operatorname{supp}u$ satisfies the distillation condition $\Pi_CZ^{v+w}|T\rangle^{\otimes t}\in\mathbb C\,\Pi_C|T\rangle^{\otimes t}$, then $X^uZ^v$ satisfies Eq.~\eqref{eq:collapse}, and every $F$ with $E(F)=X^uZ^v$ is harmless. (ii)~If every Z pattern of weight below $d$ satisfies the distillation condition, then every Pauli operator $E$ with $|E|_S<d$ satisfies Eq.~\eqref{eq:collapse}, and $C$ has distance at least $d$.
\end{lemma}

\begin{proof}[Proof of Lemma~\ref{lem:shadow}]
(i)~On one qubit, $X|T\rangle=(|T\rangle+iZ|T\rangle)/\sqrt2$, $XZ|T\rangle=-(i|T\rangle+Z|T\rangle)/\sqrt2$ and $Y=iXZ$. Up to a phase, therefore, $X^uZ^v|T\rangle^{\otimes t}=\sum_w\gamma_wZ^{v+w}|T\rangle^{\otimes t}$, summed over $w$ with $\operatorname{supp}w\subseteq\operatorname{supp}u$ and with $|\gamma_w|=2^{-|u|/2}$. If each projected term lies on the line $\mathbb C\,\Pi_C|T\rangle^{\otimes t}$, so does their sum, and Theorem~\ref{thm:main} applies. (ii)~Multiplying $E$ by an element of $\langle iI,S_C\rangle$ changes $\Pi_CE|T\rangle^{\otimes t}$ only by a phase. We may therefore take $E=X^uZ^v$ acting on fewer than $d$ qubits. Every pattern $v+w$ in~(i) is supported on these qubits, so it has weight below $d$, and~(i) applies. For the distance, let $E$ commute with $S_C$ and have $|E|_S<d$. Then $E$ commutes with $\Pi_C$, and $E\Pi_C|T\rangle^{\otimes t}=\Pi_CE|T\rangle^{\otimes t}$ is a nonzero multiple of $\Pi_C|T\rangle^{\otimes t}$ because $E$ is unitary. Through the isometry $V$, $E$ acts on $C$ as a Pauli operator $P$ on $r$ qubits up to a phase, and $|\tau\rangle$ is an eigenvector of $P$. A Hermitian multiple $P'$ of $P$ has eigenvalues $\pm1$, so $P'$ or $-P'$ fixes $|\tau\rangle$. Since $\nu(\tau)=r$, $P'=\pm I$. Hence $E\in\langle iI,S_C\rangle$, and every logical operator outside $\langle iI,S_C\rangle$ acts on at least $d$ qubits.
\end{proof}

The smallest instance of~(i) is a single check projecting onto $|T\rangle$ [Fig.~\ref{fig:small}(a)], which rejects $Z|T\rangle$ and projects $X|T\rangle$ and $Y|T\rangle$ onto $|T\rangle$ with probability $1/2$. The converse of~(i) can fail through cancellation, as in the two-qubit code with stabilizer $X_1X_2$, where $E=X_1X_2$ is harmless even though $Z_1$ and $Z_2$ in its expansion are harmful.

\subsection{Other gates and faults that are not Pauli operators}\label{sec:othergates}

Writing $|\Phi\rangle=(|00\rangle+|11\rangle)/\sqrt2$, the proof of Theorem~\ref{thm:main} extends to arbitrary non-Clifford unitaries through their teleportation gadgets.

\begin{proposition}\label{prop:gates}
Theorem~\ref{thm:main} holds for non-adaptive, fully postselected circuits with arbitrary non-Clifford unitaries $G_1,\dots,G_m$ in place of T gates, if $|T\rangle^{\otimes t}$ is replaced by the product $|\chi\rangle=\bigotimes_j|\chi_j\rangle$ of their resource states, where $|\chi_j\rangle=G_j|{+}\rangle^{\otimes k}$ if $G_j$ is diagonal on $k$ qubits, and the Choi state $|\chi_j\rangle=(I\otimes G_j)|\Phi\rangle^{\otimes k}$ otherwise. A Z fault immediately before or after a diagonal gate acts as Z on the corresponding qubit of its resource state.
\end{proposition}

\begin{proof}[Proof of Proposition~\ref{prop:gates}]
For a diagonal gate $G$ on qubits $q_1,\dots,q_k$, write $G|{+}\rangle^{\otimes k}=2^{-k/2}\sum_yG(y)|y\rangle$. Then $2^{k/2}\langle0|^{\otimes k}_a\prod_i\mathrm{CX}_{q_i\to a_i}|x\rangle|\chi_j\rangle=G(x)|x\rangle$, generalizing Eq.~\eqref{eq:gadget}. For any unitary $G$ on $k$ qubits, $2^k\langle\Phi|^{\otimes k}_{q,a}\,|\psi\rangle_q(I_a\otimes G_b)|\Phi\rangle^{\otimes k}_{a,b}=G|\psi\rangle_b$, where $\langle\Phi|_{q,a}=\langle00|_{q,a}H_q\mathrm{CX}_{q\to a}$ is a Clifford operation followed by removals and an ideal swap returns the output to $q$. In both cases the skeleton is Clifford, and the steps \emph{Skeleton}, \emph{Signs}, \emph{Faults} and \emph{Collapse} of the proof of Theorem~\ref{thm:main} apply with $|\chi\rangle$ in place of $|T\rangle^{\otimes t}$. For Z faults on a diagonal gate, $Z_{q_i}$ commutes with $G$ and satisfies $Z_{q_i}\langle0|_{a_i}\mathrm{CX}_{q_i\to a_i}=\langle0|_{a_i}\mathrm{CX}_{q_i\to a_i}Z_{a_i}$.
\end{proof}

By linearity of the skeleton map, $M_F=M\tilde E(F)$ for $D(F)=0$, where $\tilde E(F)$ includes the phase $\zeta''$ of the proof. A non-Pauli fault with Kraus operator $\sum_P\alpha_PP$, such as a coherent rotation, is therefore harmless if and only if $\Pi_C\sum_{P:\,D(P)=0}\alpha_P\tilde E(P)|\chi\rangle\in\mathbb C\,\Pi_C|\chi\rangle$, and similarly for multiple faults, although our fault distances and leading terms assume stochastic Pauli noise.

\section{The exact test}\label{sm:gauss}

\begin{definition}\label{def:dc}
A correct circuit for $|T\rangle$ has \emph{deterministic checks} if there are a subspace $\Sigma\subseteq\F^t$ and $\ell\in\F^t$ such that $Z^e\in\pm S_C$ for $e\in\Sigma$, $Z^\ell$ commutes with $S_C$ and $VZ^\ell V^\dagger$ acts on the output as $Z$, and $\Pi_CZ^e|T\rangle^{\otimes t}=0$ for every $e\notin\Sigma\cup(\ell+\Sigma)$.
\end{definition}

Here $\Sigma$ collects the Z patterns that are stabilizers of $C$ up to sign and $Z^\ell$ is logical Z, so every Z error on the T gates is a stabilizer, logical Z, or rejected with certainty, unlike in code-based distillation where syndromes are random. For $e\in\Sigma$ we write $Z^e=\xi(e)g_e$ with $g_e\in S_C$, defining a character $\xi$ of $\Sigma$ that we extend to a character $\chi$ of $\F^t$.

\begin{theorem}[Deterministic checks]\label{thm:dcauto}
Let a correct non-adaptive circuit for $|T\rangle$, ending in the ideal decoder of a CSS code that unencodes Z-type logical operators to $Z$, accept its fault-free run with certainty. Suppose that its Clifford operations carry every Z error immediately after a T gate to a Pauli operator that acts as $I$ or $Z$ on the qubit of every later T gate and on the data at the decoder, and that some Z errors after T gates reach the decoder as a logical Z operator, as Z errors after the last T layer on a logical support do. Then the circuit has deterministic checks.
\end{theorem}

\begin{proof}[Proof of Theorem~\ref{thm:dcauto}]
For $e\in\F^t$ let $F_e$ place Z faults immediately after the T gates in $e$. By Theorem~\ref{thm:main} and its proof, $D(F_e)=0$, $E(F_e)=Z^e$, and $K_{F_e}|0\rangle=2^{t/2}c\,V\Pi_CZ^e|T\rangle^{\otimes t}$. Propagating these faults to the end of the circuit, they commute with every later T gate by hypothesis, Clifford gates conjugate them, and projections that anticommute with the current operator flip sign. Let $\sigma(e)$ mark the flipped projections, which is linear in $e$. If $\sigma(e)\neq0$, let $(I+sQ)/2$ be the first flipped projection. Because the circuit accepts with certainty, the fault-free state $|\psi\rangle$ before it satisfies $Q|\psi\rangle=s|\psi\rangle$, so $(I+sQ)P|\psi\rangle=P(I-sQ)|\psi\rangle=0$, giving $K_{F_e}|0\rangle=0$ and $\Pi_CZ^e|T\rangle^{\otimes t}=0$. If $\sigma(e)=0$, the same propagation in the skeleton gives $MZ^e=\pm P_{\mathrm{out}}M$, where $P_{\mathrm{out}}$ is the decoded image of the Z-type data error $Z^b$. No decoder projection flips sign, so $b$ commutes with the X stabilizers and $P_{\mathrm{out}}=Z^{\lambda(e)}$ with $\lambda(e)\in\{0,1\}$ linear in $e$. For $\lambda(e)=0$, $\Pi_CZ^e=\pm\Pi_C$ and $Z^e\in\pm S_C$, while for $\lambda(e)=1$, $\Pi_CZ^e=\pm V^{-1}ZV\Pi_C$, so $Z^e$ commutes with $S_C$ and acts on the output as $Z$. Thus Definition~\ref{def:dc} holds with $\Sigma=\{e:\sigma(e)=0,\ \lambda(e)=0\}$ and any $\ell$ with $\sigma(\ell)=0$ and $\lambda(\ell)=1$.
\end{proof}

Our checks satisfy the hypotheses of Theorem~\ref{thm:dcauto}, whereas for the published circuits of Refs.~\cite{cultivation,googlecultivation} we verify Definition~\ref{def:dc} directly (Appendix~\ref{sm:algorithm}).

\begin{theorem}[Exact test]\label{thm:gaussfull}
For a circuit with deterministic checks and a configuration with $D(F)=0$ and $E(F)=X^aZ^b$,
\begin{equation}
 \Pi_CE(F)|T\rangle^{\otimes t}=\mu\,2^{-|a|/2}\bigl(g_0+g_1Z^\ell\bigr)\Pi_C|T\rangle^{\otimes t},
 \label{eq:out}
\end{equation}
with $|\mu|=1$ and, for $c\in\{0,1\}$ and $w\leq a$ denoting $\operatorname{supp}w\subseteq\operatorname{supp}a$,
\begin{equation}
 g_c=\sum_{w\leq a,\;w+b\in c\ell+\Sigma}i^{|w|}\,\chi(w+b+c\ell).
 \label{eq:g}
\end{equation}
The configuration is accepted with probability $2^{-|a|}(|g_0|^2+|g_1|^2)$ relative to the fault-free run, its wrong output has weight $2^{-|a|}|g_1|^2$ (the bad weight), and it is harmful if and only if $g_1\neq0$.
\end{theorem}

\begin{proof}[Proof of Theorem~\ref{thm:gaussfull}]
By Eq.~\eqref{eq:split} on $\operatorname{supp}a$ and $X^aZ^b=(-1)^{|a\wedge b|}Z^bX^a$, we have $X^aZ^b|T\rangle^{\otimes t}=(-1)^{|a\wedge b|}\,2^{-|a|/2}\sum_{w\leq a}i^{|w|}Z^{b+w}|T\rangle^{\otimes t}$. Under $\Pi_C$, terms with $b+w\notin\Sigma\cup(\ell+\Sigma)$ vanish by Definition~\ref{def:dc}. For $e\in c\ell+\Sigma$, writing $Z^{e+c\ell}=\xi(e+c\ell)g$ with $g\in S_C$ and using $[Z^\ell,\Pi_C]=0$ gives $\Pi_CZ^e|T\rangle^{\otimes t}=\chi(e+c\ell)Z^{c\ell}\Pi_C|T\rangle^{\otimes t}$, proving Eq.~\eqref{eq:out} with $\mu=(-1)^{|a\wedge b|}$. Since $V$ maps $\Pi_C|T\rangle^{\otimes t}$ and $Z^\ell\Pi_C|T\rangle^{\otimes t}$ to orthogonal vectors proportional, with equal modulus, to $|T\rangle$ and $Z|T\rangle$, the acceptance and bad weight follow.
\end{proof}

\begin{proposition}\label{prop:poly}
Each $g_c$ is computed with $O(t^3)$ operations, and $|g_c|$ is zero or a power of $\sqrt2$. In particular the acceptance and the bad weight of every fault configuration are dyadic rationals.
\end{proposition}

\begin{proof}[Proof of Proposition~\ref{prop:poly}]
Gaussian elimination in $O(t^3)$ writes $\{w\leq a: w+b\in c\ell+\Sigma\}$, when nonempty, as an affine subspace $w_0+\langle k_1,\dots,k_m\rangle$ of $\F^{|a|}$. By inclusion-exclusion, $|v_0+\dots+v_m|=\sum_{S\neq\emptyset}(-2)^{|S|-1}|\bigwedge_{s\in S}v_s|$, where terms with $|S|\geq3$ vanish modulo four. Writing $\chi(w)=(-1)^{x\cdot w}$, the summand of Eq.~\eqref{eq:g} is $i^{q(z)}$ with
\begin{widetext}
\begin{equation}
 q(z)=q_0+\sum_lL_lz_l+2\sum_{l<l'}B_{ll'}z_lz_{l'},\quad L_l=|k_l|-2|w_0\wedge k_l|+2\,x\cdot k_l,\quad B_{ll'}=|k_l\wedge k_{l'}|\bmod2 ,
\end{equation}
\end{widetext}
and $q_0=|w_0|+2\,x\cdot(w_0+b+c\ell)$, all modulo four, with $x\cdot k_l$ and $x\cdot(w_0+b+c\ell)$ reduced modulo two before doubling. To eliminate $z_v$, let $N=\{l:B_{vl}=1\}$ and $y=\bigoplus_{l\in N}z_l$. Summing over $z_v$ gives $1+i^{L_v}(-1)^y$. If $L_v$ is odd, this equals $(1+i)\,i^{3y}$ or $(1-i)\,i^{y}$, and writing $y\equiv\sum_{l\in N}z_l-2\sum_{l<l',\;l,l'\in N}z_lz_{l'}\pmod4$ keeps $q$ quadratic. If $L_v$ is even, the sum is $2[y=L_v/2]$, which vanishes or doubles when $N=\emptyset$ and otherwise keeps $q$ quadratic upon substituting one variable of $N$ as the parity of the others. Each step costs $O(m^2)$ and multiplies the modulus by $\sqrt2$, $2$ or $0$, so $|g_c|$ is zero or a power of $\sqrt2$ and the probabilities are dyadic.
\end{proof}

Judging the terms of Eq.~\eqref{eq:g} individually without phase cancellation would give fault distance one for every circuit we studied, because a single X fault before a check has a component in $\ell+\Sigma$ that vanishes only through $g_1=0$.

\subsection{Evaluating circuits}\label{sm:algorithm}

To evaluate a circuit we build its skeleton in a stabilizer simulator. Each T gate on a qubit $q$ becomes $\mathrm{CX}_{q,a_j}$ followed by a Z measurement of $a_j$ (and $S^\dagger_q$ for T$^\dagger$), where $a_j$ is half of a noiseless Bell pair with a reference qubit $r_j$, and the noise channels stay in place. The state left on the references and the output qubit is the Choi state $|\Psi\rangle$ of $V\Pi_C$, with $t+1$ stabilizer generators $g_i$. With a basis of the deterministic parities of skeleton records as detectors, and the $g_i$ (completed by record parities) as observables, one detector error model gives every fault class with its probability $p_c$, detector pattern $D_c$ and observable flips $o_c$. For $D=0$, a fixed right inverse over $\F$ maps $o$ to a reference Pauli operator $R$ equal to $E(F)$ up to transposition and $\langle iI,S_C\rangle$.

Gauss--Jordan elimination on the X parts of the $g_i$ writes $|\Psi\rangle=\sum_{\alpha\in\F^\kappa}c(x_\alpha)|x_\alpha\rangle$ over $x_\alpha=x_0+\sum_i\alpha_iu_i$. For $R=X^aZ^b$, the output vector $v=(v_0,v_1)$ is
\begin{equation}
 v_y=\sum_{\alpha:\,(x_\alpha)_{\mathrm{out}}=y}c(x_\alpha)(-1)^{b\cdot x_\alpha}\omega^{|(x_\alpha)_{\mathrm{ref}}\oplus a|},
\end{equation}
and, relative to the fault-free output $v^{(0)}$, the acceptance and the bad weight are $|v|^2/|v^{(0)}|^2$ and $|v|^2/|v^{(0)}|^2-|\langle v^{(0)},v\rangle|^2/|v^{(0)}|^4$. A Z pattern $e$ multiplies the term of $x_\alpha$ by $(-1)^{e\cdot x_0}(-1)^{\alpha\cdot\beta(e)}$ with $\beta(e)=(e\cdot u_1,\dots,e\cdot u_\kappa)$, so two Walsh--Hadamard transforms of length $2^\kappa$ evaluate all Z patterns at once. They test Definition~\ref{def:dc}, which holds if only $\beta=0$ and one other value give a nonzero output. Then $\Sigma=\beta^{-1}(0)$, $\ell$ has the other value, $\chi(e)=(-1)^{e\cdot x_0}$, and Proposition~\ref{prop:poly} decides every fault class. Every circuit behind our results satisfies Definition~\ref{def:dc}, which fails only for variants with the wrong global parity of T$^\dagger$ gates. The circuits that we evaluated have up to $263$ qubits and $101$ T gates, and $\kappa$ is $5$ and $15$ for the $\delta=3$ and $\delta=5$ circuits of Ref.~\cite{cultivation} and $27$ for the $\delta=5$ circuit of Ref.~\cite{chenrp2}.

The least number of fault classes whose detector patterns cancel and whose error is harmful is found by a meet-in-the-middle search on hashed detector patterns, and every candidate is confirmed exactly. Logical error rates come from sampling the detector error model at moderate $p$ and, for rare failures, from exact sums over sets of up to three fault classes and importance sampling of larger sets, up to eight classes, with exact completion of the last pair. On the $\delta=3$ circuit of Ref.~\cite{cultivation} at $p=10^{-3}$ under uniform noise, this estimator gives $6.17\times10^{-7}$, against $6.14(23)\times10^{-7}$ from direct sampling.

A second implementation tracks the Choi state of the skeleton symbolically, without stim, and evaluates each fault class by state-vector simulation. On the $\delta=3$ circuit of Ref.~\cite{cultivation} it agrees with the first on all $2444$ single faults and on $400$ random pairs, on every rejection and to $10^{-15}$ on accepted configurations, and both find the same $96$ malignant pairs (Appendix~\ref{sm:gsj}).

\section{Cultivation and distillation}\label{sm:distill}

\begin{corollary}\label{cor:tcount}
In a noise model in which every T gate may suffer a Z fault, every accepted fault-free run of a protocol of fault distance $d$ executes at least $n_{\min}(\tau,d)$ T gates, the least number of noisy T inputs of any order-$d$ distillation protocol for $\tau$.
\end{corollary}
\begin{proof}
By Proposition~\ref{prop:branch} it suffices to consider a correct non-adaptive circuit with $t$ T gates and fault distance $d$. Z faults at T gates have $D=0$ and $E=Z^e$, so by Theorem~\ref{thm:main} every pattern with $|e|<d$ satisfies $\Pi_CZ^e|T\rangle^{\otimes t}\in\mathbb C\,\Pi_C|T\rangle^{\otimes t}$. Measuring $C$ on $t$ noisy T states, postselecting and decoding with $V$ is then an order-$d$ distillation protocol~\cite{distillationpaper}.
\end{proof}

Let the \emph{order} $d_C$ of $C$ be the least weight of a Z pattern violating the distillation condition, and let $d_{\mathrm{sp}}=\min\{\wt(F):D(F)=0,\ |E(F)|_S>\wt(F)\}$ be the \emph{spreading distance} of the circuit.

\begin{theorem}\label{thm:attain}
A correct non-adaptive circuit with code $C$, whose T gates may suffer Z faults, has fault distance $d$ with $\min(d_C,d_{\mathrm{sp}})\leq d\leq d_C$. Conversely, for every stabilizer code $C$ on $t$ qubits that encodes a target $\tau$ with $\nu(\tau)=r$ as $\Pi_C|T\rangle^{\otimes t}\neq0$, there is a correct non-adaptive circuit with $t$ T gates and an admissible ideal decoder whose code on the T gates is $C$ and whose fault distance is $d_C$.
\end{theorem}

The converse uses the following checked cat-state measurement, which Appendix~\ref{sm:strong} uses again.

\begin{lemma}[Checked cat-state measurement]\label{lem:cat}
Let a Pauli operator $g$ act on $w$ data qubits with letters $P_1,\dots,P_w$. Prepare $(|0\rangle^{\otimes w}+|1\rangle^{\otimes w})/\sqrt2$ on cat qubits $c_1,\dots,c_w$ by preparing one of them in $|{+}\rangle$, the others in $|0\rangle$, and applying CNOTs along a spanning tree from it. In each of $\theta$ rounds, measure $Z_{c_i}Z_{c_j}$ for the edges $\{i,j\}$ of a connected graph on the cat qubits and keep the outcomes $+1$. Each parity is measured with fresh ancillas in $|0\rangle$, on which the cat qubits act only as controls of CNOTs, such as one ancilla that is the target of CNOTs from $c_i$ and $c_j$. Apply controlled-$P_i$ from $c_i$ to the $i$th data qubit for every $i$. Measure every $c_i$ in the X basis and keep the runs whose outcomes have product $s$. If at most $\theta$ faults occur in this gadget and the run is kept, it acts on the data as $R\,(I+s'g)/2$ up to a nonzero scalar, with $s'=\pm s$ and a Pauli operator $R$ for which $R$ or $Rg$ acts on at most as many qubits as there are faults.
\end{lemma}

\begin{proof}[Proof of Lemma~\ref{lem:cat}]
Without faults, X outcomes $y$ with product $(-1)^{|y|}$ apply a nonzero multiple of $[I+(-1)^{|y|}g]/2$ to the data. Push every fault to the end of the gadget. Z errors on cat qubits (including back-action from Z errors on ancillas) and flips of the X measurements change the product of the X outcomes, while X errors on ancillas and flips of their measurements affect only the parity checks because cat qubits are only CNOT controls. Moving a data Pauli from before to after its controlled-$P_i$ adds at most a Z error on $c_i$. Let $x\in\F^w$ mark the X errors on the cat qubits at the controlled Paulis, which reach the data as $P^x=\bigotimes_iP_i^{x_i}$ with $P^{x+\mathbf1}=\pm gP^x$. A fault during the verification rounds or on $c_i$ before its controlled-$P_i$ changes at most one entry of $x$, and a fault in controlled-$P_i$ or on a data qubit adds a single-qubit data Pauli. If cat preparation is fault-free, $x$ starts at $\mathbf0$. Otherwise at most $\theta-1$ faults fall in the $\theta$ rounds, so some fault-free round enforces $x_i+x_j=0$ across a connected graph, leaving $x\in\{\mathbf0,\mathbf1\}$ before later faults act. Thus the data error is $R$ or $\pm Rg$ with $R$ supported on at most as many qubits as there are faults.
\end{proof}

\begin{proof}[Proof of Theorem~\ref{thm:attain}]
\emph{Bounds.} A Z pattern $e$ of weight $d_C$ violating the distillation condition gives harmful Z faults at the corresponding T gates with $D=0$ and $E=Z^e$ (Theorem~\ref{thm:main}), so $d\leq d_C$. Conversely, if $\wt(F)<\min(d_C,d_{\mathrm{sp}})$ and $D(F)=0$, then $|E(F)|_S\leq\wt(F)<d_C$, so $F$ is harmless by Lemma~\ref{lem:shadow}(ii).

\emph{Construction.} Let $g_1,\dots,g_m$ generate $S_C$ with signs $s_k$. Prepare $t$ data qubits in $|{+}\rangle$, apply T to each, measure $g_1,\dots,g_m$ in turn by Lemma~\ref{lem:cat} with $\theta=d_C-1$ and $s=s_k$ (fixing one accepted outcome string per cat gives an equivalent fully postselected circuit), and pass the data to the ideal decoder of $C$. Each decoder projection onto $g_k$ follows the measurement of $g_k$ across commuting operations and so is a skeleton detector. Without faults the measurements apply $\Pi_C$, giving a nonzero multiple of $\Pi_C|T\rangle^{\otimes t}$, and $\sqrt2\langle0|_a\mathrm{CX}_{q\to a}|{+}\rangle_q|\psi\rangle_a=|\psi\rangle_q$ identifies the skeleton code with $C$. Now let $\wt(F)<d_C$ with all cat checks passing. By Lemma~\ref{lem:cat}, the $k$th gadget acts as $R_k(I+s'_kg_k)/2$ with $R_k$ or $R_kg_k$ supported on at most the number of faults in that gadget, and every fault outside the gadgets acts as a single-qubit data Pauli (using $T|{-}\rangle=ZT|{+}\rangle$). Moving all data Paulis to the T inputs turns the action into $\prod_k(I+s_kg_k)/2\cdot\prod_k(I+s''_kg_k)/2\cdot R\,|T\rangle^{\otimes t}$, which vanishes unless $s''_k=s_k$ for all $k$ and then equals a multiple of $\Pi_CR|T\rangle^{\otimes t}$. Choosing $R_k$ or $R_kg_k$ keeps $R$ in $R\langle iI,S_C\rangle$ and gives $|R|_S\leq\wt(F)<d_C$, so $F$ is harmless by Lemma~\ref{lem:shadow}(ii) and $d=d_C$.
\end{proof}

Note that the converse relies on the ideal decoder re-measuring $S_C$, whereas a realistic final stage uses noisy operations (Appendix~\ref{sec:final}), while Corollary~\ref{cor:tcount} needs no ideal decoder.

\begin{proof}[Proof of Theorem~\ref{cor:exact}]
Corollary~\ref{cor:tcount} gives the lower bound $n_{\min}(\tau,d)$. Conversely, Theorem~1 of Ref.~\cite{distillationpaper} gives from an optimal order-$d$ distillation protocol on $n=n_{\min}(\tau,d)$ inputs a stabilizer code $C$ with $d_C\geq d$, which Theorem~\ref{thm:attain} realizes with $n$ T gates and fault distance $d_C\geq d$. By Lemma~\ref{lem:shadow}(ii), the code of every order-$d$ distillation protocol has $r$ logical qubits and distance at least $d$, so it corrects $\lfloor(d-1)/2\rfloor$ errors and has length at least $6\lfloor(d-1)/2\rfloor-1$ by Rains' bound~\cite{rains}, giving $3d-4$ for odd $d$. For even $d$, feeding one input with $I/2$ yields an order-$(d-1)$ protocol~\cite{distillationpaper}, so $n_{\min}(\tau,d)\geq n_{\min}(\tau,d-1)+1\geq3d-6$.
\end{proof}

The exact distillation T-counts of Ref.~\cite{distillationpaper} thus transfer to cultivation, including eight T gates for $|\CCZ\rangle$ at $d=2$, while Rains' bound improves $\max(2d,d+7)$ for $|T\rangle$ from $d=7$ on, extending links between distance and suppression order known for transversal-T and GF(4)-linear codes~\cite{bravyihaah,kalraprakash}.

\paragraph*{Constructions and code order.}\label{sm:codes}
Applied to the $15$-to-$1$ code~\cite{bravyikitaev} with one verification round, Theorem~\ref{thm:attain} gives a $d=3$ circuit with $15$ T gates, $24$ qubits and $310$ two-qubit gates, so $10\leq n_{\min}(T,3)\leq15$ by Ref.~\cite{distillationpaper}. In the $\delta=3$ circuit of Ref.~\cite{cultivation} ($15$ T gates), all weight-one and weight-two Z patterns collapse while $35$ of weight three do not, giving $d_C=3$, whereas $d_{\mathrm{sp}}=2$. Among undetected fault pairs, those with $|E|_S\leq4$ are all harmless, while $48$ of the $87$ pairs with $|E|_S=5$ are harmful.

\subsection{The smallest case}\label{sm:t9}

By Theorem~\ref{cor:exact}, fault distance two for $|T\rangle$ needs nine T gates. While Theorem~\ref{thm:attain} uses $16$ qubits and $128$ two-qubit gates on the nine-input code of Ref.~\cite{distillationpaper}, an integer program over the fault map finds a smaller circuit.

\begin{proposition}[computer-assisted]\label{prop:t9}
There is a circuit with nine T gates, eleven qubits and $34$ two-qubit gates, ending in an admissible ideal decoder, that prepares one T state with fault distance two.
\end{proposition}

\begin{figure*}[t]
\centering
\input{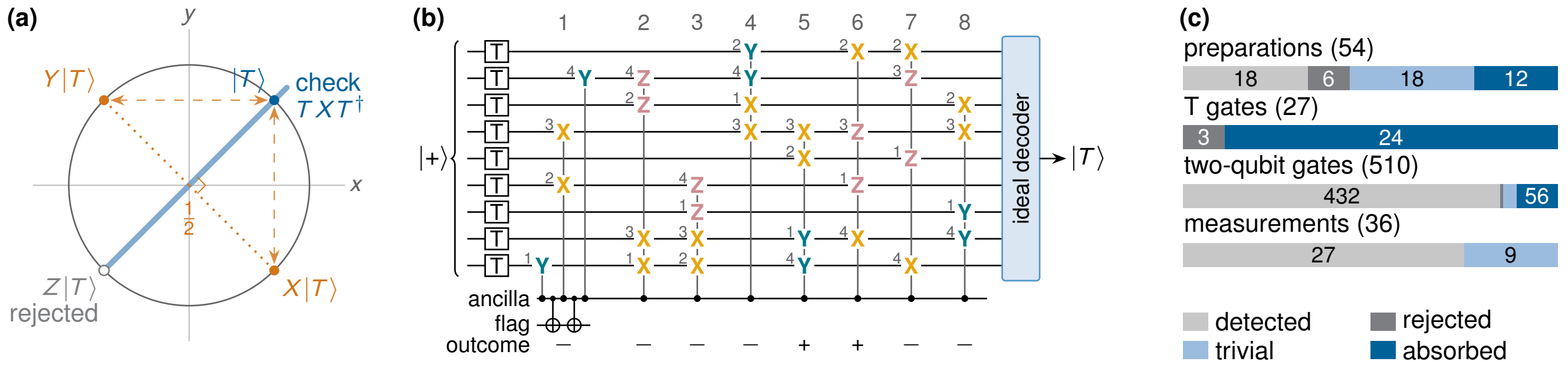}
\caption{Fault distance two with nine T gates. (a)~Equatorial Bloch-sphere view of a single check onto $|T\rangle$, which rejects $Z|T\rangle$ and projects $X|T\rangle$ and $Y|T\rangle$ onto $|T\rangle$ with probability $1/2$. (b)~Circuit of Proposition~\ref{prop:t9}. Nine qubits prepared in $|{+}\rangle$ receive T gates, followed by eight weight-four stabilizer measurements (columns 1 to 8, gate order indicated by small numbers, kept outcomes at bottom), with one flag on column 1 and an ideal decoder at the end. (c)~Breakdown of all $627$ single faults into detected, rejected, trivial (in $S_C$) and absorbed [as in (a)], where narrow two-qubit-gate segments mark $4$ rejected and $18$ trivial faults, with zero harmful faults.}
\label{fig:small}
\end{figure*}

The circuit [Fig.~\ref{fig:small}(b)] prepares nine data qubits in $|{+}\rangle$, applies T to each, and measures eight weight-four stabilizers of the distance-three nine-input distillation code~\cite{distillationpaper} with one ancilla each, postselecting on the accepted outcomes (fault-free acceptance $1/128$) before the admissible ideal decoder ($70$ locations in all). Because every weight-two input Pauli modulo stabilizers is harmful, weight-two ancilla hooks must either anticommute with an earlier measured stabilizer or trigger a flag~\cite{chaoreichardt}. A mixed-integer program over the stabilizer basis, the order of measurements and gates and the placement of flags shows that one flag on the first stabilizer suffices. This physical-level distillation~\cite{itogawa} minimizes T gates rather than cost. Under uniform noise its leading error is $53p^2$ without idle noise and $1.9\times10^3p^2$ with idle noise (which leaves the fault distance unchanged).

\begin{proof}[Proof of Proposition~\ref{prop:t9}]
Let $g_1,\dots,g_8$ be the measured stabilizers in order [Fig.~\ref{fig:small}(b)], generating the stabilizer group $S$ of the skeleton code $C$. A fault after a preparation or T gate acts as a single-qubit Pauli on the T inputs, whose Z part collapses by the single-fault certificate of Ref.~\cite{distillationpaper} and whose X part collapses by Lemma~\ref{lem:shadow}(i). A fault during the measurement of $g_k$ propagates to a data Pauli $R$, an outcome flip $\phi$ of $g_k$ and a flag flip $\psi$. It is detected by the flag if $\psi=1$, and by the decoder if $R$ anticommutes with some $g_j$ ($j<k$) or if $\phi\neq\langle R,g_k\rangle$. Otherwise $R$ commutes with $g_1,\dots,g_{k-1}$ and moves to the inputs. Exact enumeration over $\F$ partitions the $573$ measurement-circuit faults into $37$ flagged, $288$ anticommuting with an earlier $g_j$, $152$ with $\phi\neq\langle R,g_k\rangle$, and $96$ leaving $R$ of weight at most one modulo $S$, which collapses. Measurement flips and decoder-input faults fall into the same cases, and two Z faults at T gates are harmful, giving $d=2$.
\end{proof}

\paragraph*{What the proxies miss here.}
Because the circuit measures $C$ on unencoded T states, replacing T by S or Z still prepares an encoded $Q|{+}\rangle$ ($Q=T,S,Z$, in a single logical frame after writing two T gates as $ST^\dagger$) but drops the fault distance to one, with $8$ and $74$ harmful single faults. For instance, $Z_8$ is rejected by $\Pi_CZ_8|T\rangle^{\otimes9}=0$ with T gates yet accepted as a logical error with S or Z gates. Measuring each stabilizer twice with flags ($96$ two-qubit gates) satisfies the fault-tolerance condition of Ref.~\cite{simsek} with T gates while $8$ and $75$ single faults violate it with S and Z gates, answering in the negative the question of Ref.~\cite{simsek} whether non-fault-tolerant S and Z versions imply non-fault-tolerant T gates.

\section{Checks on an encoded state}\label{sm:strong}

This appendix proves the results of Sec.~\ref{sec:structure}, in their order there: the sufficient condition and the coherent family, coherent completion, the Pauli algebra of late errors and their dominance, the S proxy bound, and the correcting final stage.

\paragraph*{Setting.}
Consider a circuit starting from an ideally encoded state $|\psi\rangle$ of an $n$-qubit stabilizer code $Q$ of distance $\delta$ (stabilizer group $S$) and ending in an admissible ideal decoder of $Q$. Without faults, its data operations are $A_1,\dots,A_m$, each either a layer of single-qubit unitaries or a deterministic projection $(I+s_jP_j)/2$ ($P_j$ fixing the fault-free state with eigenvalue $s_j$) implemented by a Clifford gadget. By the proof of Theorem~\ref{thm:main}, a kept gadget run acts on the data as $R(I+s'_jP_j)/2$ up to scalar, and faults outside gadgets act as data Paulis between operations, so a kept configuration $F$ applies $K_F\propto E_mA'_mE_{m-1}\cdots A'_1E_0$ with $s'_j$ replacing $s_j$. Theorem~\ref{thm:strong} assumes that for every $F$ of $s<\delta$ faults, using $R(I+s'P)/2=s'RP(I+s'P)/2$, the supports of the $E_i$ have union $S_F$ of size at most $s$. Throughout this appendix every data qubit has a fault location immediately after each layer, and each check gadget acts after the opening layer $U^\dagger$.

\begin{proof}[Proof of Theorem~\ref{thm:strong}]
Expanding each $(I+s'_jP_j)/2$ writes $K_F$ as a sum of terms $E_mA^x_mE_{m-1}\cdots A^x_1E_0$, where each $A^x_j\in\{A_j,I,s'_jP_j\}$ is a tensor product of single-qubit unitaries. Moving $E_i$ past $A^x_m\cdots A^x_{i+1}$ conjugates it within its support, giving $O_xB_x$ with $\operatorname{supp}O_x\subseteq S_F$ and $B_x=A^x_m\cdots A^x_1$. Because the outcomes are deterministic, induction over $j$ shows that every $B_x$ maps $|\psi\rangle$ to a scalar multiple of the fault-free output $|\psi_{\mathrm{out}}\rangle\in Q$. Expanding $\sum_xc_xO_x$ into Pauli operators supported on $S_F$ ($|S_F|<\delta$), those commuting with $S$ lie in $\langle iI,S\rangle$ and act on $|\psi_{\mathrm{out}}\rangle$ as phases, while the rest anticommute with $S$ and are rejected by the decoder. Hence $F$ is harmless.
\end{proof}

\begin{theorem}[Coherent family]\label{thm:family}
In the setting of Theorem~\ref{thm:strong}, let the layers be diagonal. Let a configuration $F$ that the measurement gadgets accept leave the Pauli errors $X^{a_1}Z^{b_1},\dots,X^{a_k}Z^{b_k}$ on the data, let $A=\sum_ia_i$ and $B=\sum_ib_i$, and let $W$ be spanned by the data qubits in $\operatorname{supp}a_i$ of every error that a layer acts on after it occurs. Then every Pauli operator that reaches the decoder is, up to a phase, one of the family $X^AZ^{B+w}$ with $w\in W$, and $F$ is harmless unless the family contains a nontrivial logical operator. With S or S$^\dagger$ gates in the layers, a single member of the family reaches the decoder, the Pauli error that the S proxy tests.
\end{theorem}

\begin{proof}[Proof of Theorem~\ref{thm:family}]
Following the proof of Theorem~\ref{thm:strong}, moving $E_i=X^{a_i}Z^{b_i}$ past Pauli operators changes only its sign, while a diagonal layer $\bigotimes_q\mathrm{diag}(1,e^{i\theta_q})$ maps it by Eq.~\eqref{eq:diag} to $\prod_{q\in\operatorname{supp}a_i}(\cos\theta_q\,I-i\sin\theta_q\,Z_q)\,X^{a_i}Z^{b_i}$, a combination of $X^{a_i}Z^{b_i+w_i}$ ($w_i\leq a_i$, with $w_i=0$ if no layer follows $E_i$), and further layers preserve this form. Multiplying over $i$ writes every $O_x$ as a combination of $X^AZ^{B+w}$ ($w\in W$). The decoder rejects members anticommuting with $S$ and keeps members in $\langle iI,S\rangle$ as scalars on $|\psi_{\mathrm{out}}\rangle$, so $F$ is harmless unless the family contains a nontrivial logical operator. For $\theta_q=\pm\pi/2$, $\cos\theta_q=0$, leaving a single member. Multiplying a gadget's error by its measured operator leaves the conclusion unchanged.
\end{proof}

On a CSS code the family contains a nontrivial logical operator either when $X^A$ is a logical X and some $Z^{B+w}$ commutes with the X stabilizers (as in Lemma~\ref{lem:completion}), or when $X^A$ is an X stabilizer and some $Z^{B+w}$ is a logical Z. The second branch is absent in the proxy, where $X^A$ has a fixed Z part, whereas with T gates $w$ ranges freely over $A$ and outside Z faults can complete logical Z.

\subsection{Coherent completion}\label{sm:mechanism}

In Lemma~\ref{lem:completion}, $V_q=\mathrm{diag}(1,e^{i\theta_q})$ gives T or T$^\dagger$ gates for $\theta_q=\pm\pi/4$ and S or S$^\dagger$ gates for $\theta_q=\pm\pi/2$.

\begin{proof}[Proof of Lemma~\ref{lem:completion}]
Since $V_qXV_q^\dagger=X\,\mathrm{diag}(e^{i\theta_q},e^{-i\theta_q})=(\cos\theta_q\,I-i\sin\theta_q\,Z)X$, the error immediately after the layer is $\sum_{w\subseteq u}c_wZ^wX^u$ with $c_w=\prod_{q\in u\setminus w}\cos\theta_q\prod_{q\in w}(-i\sin\theta_q)$, and the post-layer faults turn each component into $X^{\ell\setminus u}Z^wX^u=Z^wX^\ell$. Here $X^\ell$ commutes with all stabilizers, while $Z^w$ with $w\subseteq u\subsetneq\ell$ has $|w|<\delta$ and so commutes with all X stabilizers only if $Z^w=\epsilon_wg$ for a stabilizer $g$ and sign $\epsilon_w$. The stabilizer measurements and ideal decoder remove all other components, leaving $(\sum_{w\in W_u}\epsilon_wc_w)\,X^\ell$ on the fault-free state, where $W_u=\{w\subseteq u:Z^w\in\pm S\}$. By assumption $W_u=\{\emptyset\}$, so the accepted part has amplitude $\prod_{q\in u}\cos\theta_q$ and decodes to $X_L|\tau\rangle$ (fidelity $|\langle T|X|T\rangle|^2=1/2$ for $|\tau\rangle=|T\rangle$). In a self-dual color code no Z stabilizer lies inside $\ell$, since the X stabilizer on the same support would reduce the weight of $X^\ell$. With S gates, $\cos\theta_q=0$, leaving only $Z^uX^\ell$, which has a nontrivial syndrome because $\emptyset\neq u\subsetneq\ell$ and is rejected.
\end{proof}

When $W_u$ is larger, the amplitude $\sum_{w\in W_u}\epsilon_wc_w$ can vanish. On the $[[9,1,3]]$ surface code, a boundary logical X contains a weight-two stabilizer $Z_aZ_b$, giving amplitude $\cos(\theta_a+\theta_b)$ for $u=\{a,b\}$ (accepted with certainty for T and T$^\dagger$, rejected for two T gates). On a color code with T gates, the same expansion covers the two neighboring cases seen in Appendix~\ref{sm:gsj}. First, replacing X by Y on one qubit $q\in\ell\setminus u$ requires $Z^uZ_q$ to have trivial syndrome, which holds only when $|u|=\delta-1$ and $u\cup\{q\}=\ell$, giving relative acceptance $2^{-|u|}$ and output $Y_L|\tau\rangle$ (fidelity $1/2$). Second, an error $X^\ell Z^y$ ($y\subseteq\ell$) before the last T layer leaves the two components $w=y$ and $w=\ell\setminus y$, accepted with probability $2^{1-\delta}$ and decoding to a multiple of $X_L(I+cZ_L)|\tau\rangle$ with $c=\pm i$ flipping sign with $|y|\bmod2$, giving $|T\rangle$ for one parity and an orthogonal state for the other.

\subsection{Fault tolerance relative to the code}\label{sec:relative}

For cultivation circuits on a color code of distance $\delta$, Lemma~\ref{lem:completion} bounds the fault distance directly.

\begin{corollary}\label{cor:hooks}
The fault distance of such a circuit is at most $s+\delta-|u|$ for every configuration of $s$ faults that leaves an undetected $X^u$ as in Lemma~\ref{lem:completion} immediately before the last T layer. In particular, it is at most $1+h$, where $h$ is the least number of faults that leave such an $X^u$ with $|u|=\delta-1$.
\end{corollary}

\begin{proof}
Combining the $s$ faults with post-layer X faults on $\ell\setminus u$ gives the harmful configuration of Lemma~\ref{lem:completion}.
\end{proof}

Fault tolerance relative to the code (Sec.~\ref{sec:structure}) is much weaker than ordinary ancilla fault tolerance because few subsets of weight four or more lie inside a minimum-weight logical support at larger $\delta$ ($4.6\%$ at $w=4$ for $\delta=5$, $0.04\%$ at $w=6$ for $\delta=7$, Table~\ref{tab:cover}), so lower bounds on fault-tolerant cat preparation~\cite{spidercat} need not apply to real-T checks.

\begin{table}[t]
\caption{Fraction of weight-$w$ X patterns on triangular color codes lying inside a minimum-weight logical X support (number of such supports in parentheses).}
\label{tab:cover}
\begin{ruledtabular}
\begin{tabular}{lccccc}
code & $w=2$ & 3 & 4 & 5 & 6 \\
\hline
$[[7,1,3]]$ (7) & 1 & & & & \\
$[[19,1,5]]$ (36) & 0.895 & 0.291 & 0.046 & & \\
$[[37,1,7]]$ (140) & 0.811 & 0.278 & 0.052 & 0.0060 & 0.0004 \\
\end{tabular}
\end{ruledtabular}
\end{table}

\subsection{Flat phases and late errors}\label{sec:late}

Let $Q$ be a self-dual CSS code with one logical qubit, $+1$ Z stabilizers and logical X operator $X^{\otimes n}$, as for triangular color codes. Identifying operators with their supports ($S_Z$, $\ell_Z+S_Z$, $S_X$, $\ell_X+S_X$), for a T/T$^\dagger$ layer $U$ with angles $\theta_q=\pm\pi/4$ we define
\begin{equation}
 \phi(w)=\prod_{q\in w}(-i\operatorname{sgn}\theta_q),\qquad \phi(a+b)=\phi(a)\phi(b)(-1)^{|a\cap b|}.
 \label{eq:phi}
\end{equation}

\begin{lemma}[Flat phases]\label{lem:flat}
The operator $UX^{\otimes n}U^\dagger$ maps $Q$ to itself if and only if $\phi(k)=1$ for every $k\in S_Z$. Then $\phi$ is constant on $\ell_Z+S_Z$, with a value $c\in\{i,-i\}$, and $UX^{\otimes n}U^\dagger$ acts on $Q$ as $(I+cZ_L)X_L/\sqrt2$.
\end{lemma}

\begin{proof}[Proof of Lemma~\ref{lem:flat}]
Expanding $UX^{\otimes n}U^\dagger=2^{-n/2}\sum_w\phi(w)Z^wX^{\otimes n}$, the code projector $\Pi$ satisfies $\Pi Z^w\Pi=0$ unless $w\in S_Z\cup(\ell_Z+S_Z)$, so $\Pi UX^{\otimes n}U^\dagger\Pi=2^{-n/2}(a+bZ_L)X_L\Pi$ with $a=\sum_{k\in S_Z}\phi(k)$ and $b=\sum_{k\in\ell_Z+S_Z}\phi(k)$. Unitarity on $|\psi\rangle\in Q$ with $\langle\psi|X_LZ_LX_L|\psi\rangle=0$ requires $2^{-n}(|a|^2+|b|^2)=1$. Since each sum has $2^{(n-1)/2}$ unit-modulus terms, $\phi$ is constant on each coset, with $\phi=\phi(0)=1$ on $S_Z$. Conversely, if $\phi=1$ on $S_Z$, then $|\ell_Z\cap k|$ is even for all $k\in S_Z$, so Eq.~\eqref{eq:phi} gives $\phi(\ell_Z+k)=\phi(\ell_Z)=c\in\{\pm i\}$ (as $|\ell_Z|$ is odd), yielding $(I+cZ_L)X_L/\sqrt2$ on $Q$.
\end{proof}

On a face $f$ with $t_f$ gates T$^\dagger$, $\phi(f)=(-i)^{|f|}(-1)^{t_f}=1$ holds if and only if $t_f\equiv|f|/2\pmod2$. Since faces have even overlaps, Eq.~\eqref{eq:phi} makes Ref.~\cite{chan}'s sufficient condition necessary as well, while $c$ selects $T|{+}\rangle$ or $T^\dagger|{+}\rangle$.

\begin{theorem}[Late errors]\label{thm:late}
In the setting of Theorem~\ref{thm:strong}, with the code $Q$ above, layers of T and T$^\dagger$ gates whose checks are logical, and a last check consisting of $U^\dagger$, a measurement of $X^{\otimes n}$ and $U$, let a configuration $F$ that the gadgets accept satisfy four conditions. Every error with an X part that a layer acts on occurs between this $U^\dagger$ and $U$. Every error after $U$ occurs immediately after it. Only fault-free stabilizer measurements follow $U$. Every projection before $U$ keeps its fault-free sign once the errors before it are moved past it, since otherwise $F$ is rejected. Let $X^A$ be the X part of the errors that arrive at $U$, $X^{A'}$ that of the errors after $U$, and $Z^B$ the Z part of all errors. Let $K_A$ be the set of Z stabilizers and Z logical operators supported inside $A$, and $K^0_A\subseteq K_A$ its Z stabilizers. Then $F$ is accepted if and only if $A+A'$ lies in $S_X$ or in $\ell_X+S_X$ and some $w\subseteq A$ with $|w\cap k|$ even for every $k\in K^0_A$ satisfies $B+w\in S_Z\cup(\ell_Z+S_Z)$. Its acceptance relative to the fault-free run is then $2^{-|A|}|K_A|\,|K^0_A|$. With $\mu=1$ if $A+A'\in\ell_X+S_X$ and $\lambda=1$ if $B+w\in\ell_Z+S_Z$, and $\mu=0$ or $\lambda=0$ otherwise, the following holds.
(i)~If $K_A$ contains no logical operator, the output is $Z_L^\lambda X_L^\mu|\tau\rangle$, and $F$ is harmful if and only if $(\mu,\lambda)\neq(0,0)$.
(ii)~If $K_A$ contains a Z logical operator $\ell_A$, the output is proportional to $Z_L^\lambda(I+c'Z_L)X_L^\mu|\tau\rangle$ with $c'=c\,(-1)^{|(w+A')\cap\ell_A|}$, and $F$ is harmless if and only if $\mu=1$ and $(-1)^\lambda c'=c$.
\end{theorem}

\begin{proof}[Proof of Theorem~\ref{thm:late}]
Move every Z error to immediately after $U$. If it flips the net sign of any earlier projection on its fault-free eigenstate, $F$ is rejected. Otherwise errors between $U^\dagger$ and $U$ commute in X with $X^{\otimes n}$ and move to $U$, giving $X^A|\varphi\rangle$ with $|\varphi\rangle=U^\dagger|\tau_L\rangle$. Expanding $UX^AU^\dagger$ and applying post-$U$ errors gives, up to phase, $2^{-|A|/2}\sum_{w\subseteq A}\phi(w)\,(-1)^{|A'\cap w|}\,Z^{B+w}X^{A+A'}|\tau_L\rangle$. The decoder keeps terms with $A+A'\in S_X\cup(\ell_X+S_X)$ and $B+w\in S_Z\cup(\ell_Z+S_Z)$, forming a coset $w_0+K_A$ on which $Z^{B+w}X^{A+A'}|\tau_L\rangle=\pm Z_L^{\lambda(w)}X_L^\mu|\tau_L\rangle$ with $w$-independent sign. For $k\in K^0_A$, $\phi(k)=1$ and $|k|$, $|A\cap k|$ and $|A'\cap k|$ are even, so Eq.~\eqref{eq:phi} gives $\phi(w_0+k)(-1)^{|A'\cap(w_0+k)|}=\phi(w_0)(-1)^{|A'\cap w_0|}(-1)^{|w_0\cap k|}$. Summing this character over $K^0_A$ gives $|K^0_A|$ if $|w_0\cap k|$ is even for all $k\in K^0_A$ and zero otherwise, proving~(i) when $K_A=K^0_A$. When $K_A$ also contains $\ell_A+K^0_A$, that coset carries the extra factor $\phi(\ell_A)(-1)^{|(w_0+A')\cap\ell_A|}=c'$ and flips $\lambda$, giving output proportional to $Z_L^\lambda(I+c'Z_L)X_L^\mu|\tau\rangle$ with acceptance $2\cdot2^{-|A|}|K^0_A|^2$. This has fidelity $1/2$ for $\mu=0$, while for $\mu=1$, Lemma~\ref{lem:flat} gives $(I+cZ_L)X_L|\tau\rangle=\sqrt2|\tau\rangle$ and $(I-cZ_L)X_L|\tau\rangle\propto Z_L|\tau\rangle\perp|\tau\rangle$, proving~(ii).
\end{proof}

When Z stabilizers $k$ carry signs $s_k=\pm1$, as in the $\delta=5$ check on $\mathbb{RP}^2$ of Ref.~\cite{chenrp2}, Lemma~\ref{lem:flat} and Theorem~\ref{thm:late} hold with $\phi(k)=s_k$ and $c=\phi(\ell_Z+k)s_k$, leaving the late-error criterion unchanged. Theorem~\ref{thm:late} recovers Lemma~\ref{lem:completion} and its neighboring cases while adding the even-overlap parity condition on $K^0_A$. In particular, if $s$ accepted faults leave $(A,A')$ with $B=0$, the real-T fault distance is at most $s+c(A,A')$, where $c(A,A')$ is the least weight of post-$U$ data X and Z faults making $(A,A')$ harmful.

\begin{corollary}[Face hooks]\label{cor:face}
Let $f$ be the support of a Z stabilizer of $Q$ of weight at most six that contains no other nonzero Z stabilizer and no Z logical operator, such as a face of a triangular color code, and let $f$ contain a qubit of some minimum-weight logical support. A check in which one accepted fault leaves exactly the error $X^f$ just before the last layer has real-T fault distance at most $\delta-1$, while the S proxy finds this configuration harmless and $f$ lies inside no minimum-weight logical support. On the triangular color codes of distance $\delta\leq11$ every face qualifies, and fault tolerance relative to the code is therefore necessary but not sufficient for the full distance.
\end{corollary}

\begin{proof}
For a minimum-weight logical support $L$ meeting $f$, $|L\cap f|$ is nonzero, even (since $[Z^L,X^f]=0$) and at most $|f|/2$ (since $|L+f|=|L|+|f|-2|L\cap f|\geq\delta$), so $|L\cap f|=2$ and $f\not\subseteq L$. Placing post-$U$ Z faults on the $\delta-2$ qubits of $L\setminus f$ gives $A=f$, $A'=0$, $K^0_A=K_A=\{0,f\}$ and $w=L\cap f$ meeting $f$ evenly with $B+w=L$, so case~(i) of Theorem~\ref{thm:late} gives $(\mu,\lambda)=(0,1)$, accepted with probability $2^{2-|f|}$ as $Z_L|\tau\rangle$. With S gates, $X^f$ becomes the stabilizer $\pm X^fZ^f$ and $B$ is rejected. Integer programs verify that every face of the triangular color codes up to $\delta=11$ meets some minimum-weight logical support.
\end{proof}

Conversely, if post-$U$ Z faults meet $f$ oddly, $w$ meets $f$ oddly and the two terms $w$ and $f+w$ cancel, whereas Theorem~\ref{thm:family} keeps both and errs on the safe side.

\begin{proposition}\label{prop:pairs}
On the $[[19,1,5]]$ color code, every pair of data qubits that avoids the qubit shared by the three hexagons lies in a minimum-weight logical support. A check in which each of two single accepted faults leaves X just before the last layer on its own pair of data qubits, the two pairs being disjoint, therefore has real-T fault distance at most four. This holds for every GHZ fold whose tree has two disjoint pairs of a node and its last-unfolded leaf child, and for every check with a cat grown along a path from a root with at least two sites on each side, not verified and read out without undoing its growth, whatever the placement of the data along the path. A check with a cat grown in this way from one end also has real-T fault distance at most four, for every placement.
\end{proposition}

\begin{proof}
Direct enumeration of the $36$ minimum-weight logical supports shows that all $18$ uncovered pairs contain the central three-hexagon qubit. One of two disjoint pairs avoids that qubit and lies in some $\ell$, giving four harmful faults by Lemma~\ref{lem:completion}. Such disjoint pairs arise from node-leaf unfold CNOTs in folds (Proposition~\ref{prop:fold3}) and from the penultimate sites at both ends of a middle-grown unverified path cat. For an end-grown path cat without un-growing, a single fault on any site spreads X to a path suffix $S$ (equivalent modulo $X^{\otimes n}$ to $\bar S$), and dynamic programming shows that subsets with $\min(c(S,0),c(\bar S,0))\geq4$ admit no nested chain $|S_m|=m$ past $m=8$, so some suffix has $c\leq3$ and four faults are harmful.
\end{proof}

\subsection{Errors before the last check}\label{sec:dominance}

\begin{lemma}[Check filtering]\label{lem:filter}
Let a check consist of a layer $U$ of T and T$^\dagger$ gates, applied as $U^\dagger$, a projection onto $X^{\otimes n}=s$ whose outcome is fixed without faults, and $U$. If the data enter the check in the state $Z^bX^a|\psi\rangle$ with $a\neq\emptyset$, where $|\psi\rangle$ is the fault-free state, and no fault occurs inside the check, they leave it in the state $\tfrac12\bigl(Z^bX^a+(-1)^{|b|}\phi(a)\,Z^{b+a}X^a\bigr)|\psi\rangle$, with $\phi$ of Eq.~\eqref{eq:phi}.
\end{lemma}

\begin{proof}[Proof of Lemma~\ref{lem:filter}]
Write $|\psi\rangle=U|\varphi\rangle$ with $X^{\otimes n}|\varphi\rangle=s|\varphi\rangle$. Since $U^\dagger Z^bX^aU=2^{-|a|/2}\sum_{v\subseteq a}\bar\phi(v)Z^{b+v}X^a$, the projection onto $X^{\otimes n}=s$ keeps terms with $|b+v|$ even. Applying $U$ gives $Z^{b+u}X^a$ with coefficient $2^{-|a|}\sum_v\bar\phi(v)\phi(u+v)=2^{-|a|}\phi(u)\sum_v(-1)^{|u\cap v|}$ summed over $v\subseteq a$ with $|v|\equiv|b|\pmod2$, which equals $1/2$ for $u=\emptyset$, $(-1)^{|b|}\phi(a)/2$ for $u=a$, and $0$ otherwise (for $a=\emptyset$, the check passes $Z^b|\psi\rangle$ if $|b|$ is even and rejects otherwise).
\end{proof}

Passing a complete check thus reduces the $2^{|a|}$ Z patterns of Theorem~\ref{thm:family} to just $u=\emptyset$ and $u=a$, so the partial Z patterns needed by Lemma~\ref{lem:completion} and Corollary~\ref{cor:face} must originate inside a check. Moreover, Theorem~\ref{thm:late} extends unchanged to noisy post-$U$ faults and stabilizer gadgets acting as $R(I+s'g)/2$, with the added acceptance condition that intermediate projections keep the surviving coset $w_0+K_A$ (since its members share the same X part and differ by $Z^k$, $k\in K_A$, commuting with all stabilizers).

\begin{theorem}[Late errors dominate]\label{thm:dominance}
In the setting of Theorem~\ref{thm:late}, with faults allowed at every location and a single check whose gadget acts only after its $U^\dagger$, suppose that
(i)~for every configuration $F_e$ of $k<\delta$ faults before the $U^\dagger$ of the check that the gadgets accept, which leaves $Z^bX^a$ on the data, some such configuration $G$ that leaves only a Z error $Z^c$ satisfies $|G|+\wt_S(Z^{b+c}X^a)\leq k$, where $\wt_S(P)$ is the least number of qubits of a Pauli operator equal to $P$ up to stabilizers,
(ii)~every detector is a sum of detectors whose records all lie before, or all after, this $U^\dagger$, and
(iii)~every data qubit has a fault location between its gate of this $U^\dagger$ and the gadget of the check.
If a configuration $F$ of fewer than $\delta$ faults is harmful, then some late configuration with at most as many faults is harmful. This configuration keeps the faults of $F$ from the $U^\dagger$ on, replaces the earlier ones by a configuration $G$ that leaves only Z errors, and adds single-qubit faults inside the check. The real-T fault distance is therefore the smaller of $\delta$ and the least weight of a harmful late configuration, which Theorem~\ref{thm:late} decides by Pauli algebra.
\end{theorem}

\begin{proof}[Proof of Theorem~\ref{thm:dominance}]
Let $F_e$ be the $k$ faults of $F$ before $U^\dagger$, leaving $P|\psi\rangle$ ($P=Z^bX^a$) at the check input, where we choose $Z^{b+c}X^a$ (up to stabilizers fixing $|\psi\rangle$) to have weight at most $\wt_S(Z^{b+c}X^a)$ up to Z stabilizers for the $G$ of~(i). With check-gadget error $R$ and $|\varphi\rangle=U^\dagger|\psi\rangle$, the expansion in Lemma~\ref{lem:filter} gives
\begin{widetext}
\begin{equation*}
 UR\,\Pi\,U^\dagger P|\psi\rangle=\sum_v\bar c_v\,UR\,Z^{b+v}X^a|\varphi\rangle,\qquad\bar c_v=2^{-|a|/2}\bar\phi(v),
\end{equation*}
\end{widetext}
summed over $v\subseteq a$ of fixed parity. Each term, followed by the post-$U$ part of $F$, equals the run of the late configuration $L_v$ formed by $G$ (leaving $Z^c$, which commutes with $U^\dagger$), single-qubit faults creating $Q_v=Z^{b+c+v}X^a$ up to a Z stabilizer just before the check gadget via~(iii), and the faults and records of $F$ from $U^\dagger$ on, with all detectors vanishing by~(ii). Since the decoded output of $F$ is a nonzero-coefficient sum of those of $L_v$, some $L_v$ is harmful, of weight at most $|G|+\wt_S(Z^{b+c}X^a)+\wt(F\setminus F_e)\leq\wt(F)$.
\end{proof}

Conditions~(i) and~(ii) hold whenever the circuit starts from an ideally encoded state with either no pre-check stabilizer rounds (as in Proposition~\ref{prop:nn}) or verified-cat stabilizer rounds (as in Corollary~\ref{cor:shor}, and even when flagged rounds leave pure Z hooks via $G=F_e$), whereas for single-flag pre-check rounds the agreement between the late bound and exact certificates is empirical.

\subsection{The S proxy as an upper bound}\label{sec:proxybound}

On a self-dual CSS code $Q$, replacing T/T$^\dagger$ by S/S$^\dagger$ ($\theta_q=\pm\pi/2$) turns $U_SX^{\otimes n}U_S^\dagger\propto Z^{\otimes n}X^{\otimes n}$ into logical Y up to sign, with deterministic $+1$ eigenstate $\tau_S$.

\begin{lemma}[T against S]\label{lem:sproxy}
A late configuration with the patterns $A$, $A'$ and $B$ of Theorem~\ref{thm:late} is accepted by the S proxy if and only if $A+A'\in S_X\cup(\ell_X+S_X)$ and $B+A\in S_Z\cup(\ell_Z+S_Z)$. It is then harmful for the proxy if and only if exactly one of $\mu$ and $\lambda$ is one, with $\lambda$ taken for $w=A$. Every late configuration that is harmful with S gates is harmful with T gates.
\end{lemma}

\begin{proof}[Proof of Lemma~\ref{lem:sproxy}]
For $\theta_q=\pm\pi/2$, $V_qXV_q^\dagger=\mp iZX$ leaves only $w=A$, delivering $Z^{B+A}X^{A+A'}|\tau_{S,L}\rangle$ to the decoder, which accepts under the stated coset conditions and outputs $Z_L^\lambda X_L^\mu|\tau_S\rangle$. Since only logical Y fixes $\tau_S$, this is harmful if and only if $(\mu,\lambda)\in\{(1,0),(0,1)\}$. For any $k\in K^0_A$, $[Z^k,X^{\otimes n}]=0$ implies $|A\cap k|=|k|$ is even, so $w=A$ satisfies the parity condition of Theorem~\ref{thm:late} and is accepted with T gates with the same $(\mu,\lambda)$. It is then harmful in case~(i) because $(\mu,\lambda)\neq(0,0)$, and in case~(ii) either because $\mu=0$ or because $(\mu,\lambda)=(1,0)$ implies $\{X^{A+A'},Z^{\ell_A}\}=0$, so $|(A+A')\cap\ell_A|$ is odd and $c'=-c\neq(-1)^\lambda c$.
\end{proof}

By self-duality, $S_X=S_Z$ and $\ell_X+S_X=\ell_Z+S_Z$. When $\mu+\lambda=1$, one of $A+A'$ and $B+A$ is a stabilizer and the other logical, so $A'+B$ is logical and $|A'\cup B|\geq|A'+B|\geq\delta$, eliminating $A$. Noisy post-$U$ projections preserving the $w=A$ member likewise preserve its entire T-family coset.

\begin{proof}[Proof of Theorem~\ref{thm:proxybound}]
Let $d_S$ be the proxy fault distance. If $d_S=\delta$, Proposition~\ref{prop:singleton} gives $d\leq\delta$. If $d_S<\delta$, take a harmful proxy configuration $F$ of weight $d_S$, with $k$ pre-$U_S^\dagger$ faults leaving $Z^bX^a$. By condition~(i) of Theorem~\ref{thm:dominance}, choose an accepted pre-$U_S^\dagger$ configuration $G$ leaving $Z^c$ with $|G|+\wt_S(Z^{b+c}X^a)\leq k$, let $R=Z^{b+c}X^ag$ ($g\in S$) have weight $\wt_S(Z^{b+c}X^a)$, and place single-qubit faults creating $R'=U_S^\dagger RU_S$ just before the check gadget via~(iii). Combining $G$, those faults and the post-$U_S^\dagger$ part of $F$ gives an accepted late configuration $L$ of weight at most $d_S$ that reaches the check gadget in the same state $\propto U_S^\dagger Z^bX^a|\psi\rangle$ as $F$, is harmful for the proxy, and by Lemma~\ref{lem:sproxy} is harmful with T gates, so $d\leq d_S$.
\end{proof}

\begin{proof}[Proof of Proposition~\ref{prop:blind}]
If $d_S<\delta$, the proof of Theorem~\ref{thm:proxybound} yields a harmful late proxy configuration $L$ of weight at most $d_S$ built from Z-only pre-check faults, single-qubit faults inside the check and post-$U^\dagger$ faults. By hypothesis each fault of $L$ puts Z on at most one data qubit and each post-$U$ fault acts on at most one data qubit, giving $|A'\cup B|\leq d_S<\delta$, contradicting $|A'\cup B|\geq\delta$ from Lemma~\ref{lem:sproxy}, while Proposition~\ref{prop:singleton} gives $d_S\leq\delta$.
\end{proof}

Cat checks without stabilizer rounds satisfy this hypothesis because cat qubits control CNOTs to data, preventing Z errors from spreading between data qubits.

\subsection{Certificates without T gates}\label{sec:certificates}

Theorems~\ref{thm:late} and~\ref{thm:dominance} turn the certification of a single check into a combinatorial problem without T gates. Its binary variables select single faults from the $U^\dagger$ of the check on, and single-qubit X and Z faults on the data after $U$. Its constraints make the detectors of the gadgets cancel and impose the coset conditions of Theorem~\ref{thm:late} in its three branches: $\mu=1$, or $\mu=0$ and $\lambda=1$, or $\ell_A\subseteq A$. The parity condition is omitted, so that every harmful late configuration is feasible. We solve the problem as a mixed-integer program, test each solution exactly by Theorem~\ref{thm:late} and cut it off if it is harmless. By Theorem~\ref{thm:dominance} and Proposition~\ref{prop:singleton}, infeasibility below $\delta$ certifies $d=\delta$. Because errors $X^aZ^b$ from before $U_S^\dagger$ act as $X^aZ^{a+b}$ inside the check of the S proxy, infeasibility without cuts also certifies proxy fault distance $\delta$.

A floating-point solver (Gurobi) establishes infeasibility for the ladders at $\delta=7$ and $9$ and for the verified cats without stabilizer rounds at $\delta=7$ and $9$. A second solver (HiGHS) confirms the $\delta=9$ ladder and the $\delta=7$ verified cat, and an exact search with Gauss sums, which uses neither the integer program nor Theorem~\ref{thm:dominance}, excludes every harmful configuration of up to six faults in the $\delta=7$ verified cat without stabilizer rounds. A simpler bound, which propagates the single faults of the check and minimizes $s+c(A,A')$ over combinations whose detectors cancel, reproduces the exact fault distance of every fold and unverified cat at $\delta=3$ and $5$, of the central fold tree at $\delta=7$, and of the grid checks with a single end-parity verification described next.

Whether a single verification suffices at every $\delta$ is open, because a constant number of faults can leave X on a tail of the path, which places such checks outside Theorem~\ref{thm:strong}. At $\delta=5$, a grid check with $2n+1=39$ qubits whose cat growth is reversed and verified only by the parity of its ends, without rungs or diagonals, keeps the full fault distance in seven of twelve placements of the data.

\subsection{A correcting final stage}\label{sec:final}

An ideal correcting decoder of an $[[m,r,\delta']]$ code accepts every run, applying a minimum-weight correction for the measured syndrome before unencoding.

\begin{proof}[Proof of Proposition~\ref{prop:final}]
\emph{Upper bound.} Partition a weight-$\delta'$ logical Pauli $P$ into $R$ of size $\lfloor\delta'/2\rfloor+1$ and $\bar R$ of size $\lceil\delta'/2\rceil-1$. Placing $P_R$ just before the decoder produces the syndrome of $P_{\bar R}$, so the minimum-weight correction $C$ has weight at most $|\bar R|$. Then $CP_{\bar R}$ has trivial syndrome and weight at most $2\lceil\delta'/2\rceil-2<\delta'$, hence lies in $\langle iI,S\rangle$, so $CP_R$ applies a nontrivial logical Pauli to $\tau$ ($\nu(\tau)=r$) with $\lfloor\delta'/2\rfloor+1$ faults.

\emph{Lower bound.} For $s\leq t=\lfloor(\delta-1)/2\rfloor$ faults, the proof of Theorem~\ref{thm:strong} delivers $\sum_P\beta_PP|\psi_{\mathrm{out}}\rangle$ with $|\operatorname{supp}P|\leq s$. For any observed syndrome $\sigma$ occurring in the sum, the correction $C_\sigma$ has weight at most $t$, so $C_\sigma P$ has trivial syndrome and weight at most $2t<\delta$, lying in $\langle iI,S\rangle$ and returning $|\psi_{\mathrm{out}}\rangle$.
\end{proof}

Reaching fault distance $d$ thus requires keeping postselection until the code grows to distance at least $2d-2$, as in the confidence-postselected escape stage of Ref.~\cite{cultivation}. While Theorems~\ref{thm:main} and~\ref{thm:gauss} apply unchanged to a noisy Clifford growth stage, certifying noisy growth remains open because its random measurement outcomes lie outside Theorem~\ref{thm:strong} and later noisy projections can pass coherent error components that an ideal decoder rejects.

\section{GHZ folds and the published circuits}\label{sm:gsj}

This appendix applies Appendix~\ref{sm:strong} to GHZ folds in general and then certifies the published circuits.

\subsection{GHZ folds at every distance}\label{sec:folds}

In the GHZ-fold double checks of Ref.~\cite{cultivation}, $h=1$ at $\delta=3$ and $h=2$ at $\delta=5$, and every malignant configuration has the form of Lemma~\ref{lem:completion} or its neighboring cases with relative acceptance $2^{-(\delta-1)}$ (Table~\ref{tab:full}), attaining Corollary~\ref{cor:hooks}. Consider in general a check round that pairs each data qubit with a partner via $\mathrm{CX}(\text{partner}\to\text{data})$, folds over a spanning tree on one representative per pair, measures and resets the root, unfolds, unpairs and applies the last T layer, followed only by stabilizer measurements and the ideal decoder.

\begin{proposition}\label{prop:fold3}
Let the code be a CSS code of distance $\delta$ in which no Z stabilizer lies inside a minimum-weight logical X support. If $k<\delta/2$ disjoint pairs, each formed by the data qubits of a node of the tree and of its last-unfolded child, which is a leaf, lie in one minimum-weight logical X support, then the real-T fault distance of the check is at most $\delta-k$. On the $[[7,1,3]]$ code, every such check has real-T fault distance at most two, for every tree, root, choice of representatives and order of the CNOTs.
\end{proposition}

\begin{proof}
An X fault on a node $v$ just before its last unfold CNOT to a leaf child $c$ spreads only to $c$, and unpairing leaves X on the data qubits of $v$ and $c$ regardless of representatives, unseen by the earlier root measurement or the partner X measurements. Thus $k$ such faults leave $X^u$ with $|u|=2k<\delta$ before the last T layer, and $\delta-2k$ post-layer data faults complete $\ell$, giving $\delta-k$ harmful faults by Lemma~\ref{lem:completion}. Every tree on at least two nodes has a non-leaf node whose children are all leaves, and on $[[7,1,3]]$ every pair of qubits lies on one of the seven Fano lines, so $k=1$ always applies.
\end{proof}

On $200$ random labeled trees on $[[7,1,3]]$, with random roots and representatives, the certified fault distance is always two. At larger distance~\cite{chan} we built GHZ-fold checks between flagged stabilizer rounds from an ideally encoded input ($T^\dagger|{+}\rangle$ at $\delta=3,7$ and $T|{+}\rangle$ at $\delta=5$, as Definition~\ref{def:dc} requires). Each check prepares $n$ partners in $|{+}\rangle$, applies $U^\dagger$ and the pairing CNOTs, folds and unfolds over a spanning tree of face-sharing qubits (unrestricted of depth one to three, or binary of depth five to seven), unpairs, applies $U$ and measures the partners in X. Every tree that we built, three each at $\delta=3$ and $5$ and six at $\delta=7$ ($t=75$), has real-T fault distance $2$, $3$ and $4$, against proxy fault distances $3$, $5$ and $7$ (Fig.~\ref{fig:fd}). At $\delta=7$ the malignant configurations have relative acceptance $2^{-10}$ to $2^{-4}$, and the leading terms range from $34\,p^4$ to $4.4\times10^3p^4$ (Table~\ref{tab:full}).

\paragraph*{Bounds from the fold tree.}
Propagating single X and $X\otimes X$ check-round faults on the tree and minimizing $s+\delta-|u|$ over combinations leaving $X^u$ ($\emptyset\neq u\subsetneq\ell$) bounds the fault distance without simulation. This tree bound matches $11$ of the $12$ trees above (giving $5$ on the central $\delta=7$ tree, where Theorem~\ref{thm:late} gives $4$) and reproduces $3$ and $4$ for the unverified cat at $\delta=5,7$. Across seven fold trees at $\delta=9,11,13$ ($[[61,1,9]]$, $[[91,1,11]]$, $[[127,1,13]]$), the bound ranges over $4$ to $6$, $6$ to $8$, and $7$ to $10$. In particular, two binary trees at $\delta=9$ have upper bounds $5$ and $4$, and without stabilizer rounds the integer program of Appendix~\ref{sec:certificates} certifies their exact fault distances as $5$ and $4$.

\subsection{Color-code cultivation}

\paragraph*{Distance three.}
Joining the two published stim chunks~\cite{gsjcode} ($15$ qubits, $15$ T gates, $81$ CNOTs), replacing S and S\_DAG by T and T$^\dagger$, and keeping the final stabilizer measurements as an admissible ideal decoder ($4$ of $20$ projections entering $C$ rather than $D$) gives $0$ harmful single faults and $48$ harmful pairs out of $966{,}391$ without idle noise ($d_C=3$ and $d_{\mathrm{sp}}=2$, Appendix~\ref{sm:codes}), whereas the S proxy has none [Fig.~\ref{fig:gsj}(a)]. These $48$ pairs form $9$ pairs of fault classes (all combining a fold X/Y fault with a final-layer data X/Y fault), or $12$ classes when grouped by sign flips and final Pauli as in Ref.~\cite{chan} ($4$ of fidelity $0$ and $8$ of fidelity $1/2$). Under the uniform depolarizing model of Ref.~\cite{gsjcode} with idle noise between preparation and measurement ($436$ locations, $299$ idle), $96$ pairs are harmful, each with relative acceptance $1/4$ and infidelity $1/2$ or $1$. Summing $\epsilon=\sum_Fw_Fh_F/\sum_Fw_Fa_F$ gives $\epsilon_T=32p^2/75+O(p^3)$ and $\epsilon_S=O(p^3)$, matching the exact series $32p^2/75+(2009039/4500)x^3$ ($x=p/(1-p)$) of Wan and Zapirain~\cite{wanzapirain} on the independently compiled circuit of Ref.~\cite{soft} ($\epsilon_T=8.98\times10^{-7}$ at $p=10^{-3}$). Weighted fits of $(\epsilon_T-\epsilon_S)/p^2=a+bp$ to the state-vector data of Ref.~\cite{gsjcode} give $a=0.42\pm0.06$ over all points [Fig.~\ref{fig:gsj}(b)] and $a=0.44\pm0.08$ for $p\leq5\times10^{-3}$ against $32/75\approx0.427$, and sampling the detector error model of the published circuit with unitary injection gives $8.94(33)\times10^{-7}$ at $p=10^{-3}$ against $(9.7\pm0.5)\times10^{-7}$ in Ref.~\cite{gsjcode}.

\paragraph*{Distance five.}
The $\delta=5$ circuit ($42$ qubits, with $\delta=3$ and $\delta=5$ double checks and code growth between them) requires two real-T conditions. First, by Lemma~\ref{lem:flat} (extending Ref.~\cite{chan}), $\bigotimes_qU_qXU_q^\dagger$ with $U_q\in\{T,T^\dagger\}$ is logical on the $19$-qubit color code if and only if each face $f$ carries $t_f\equiv|f|/2\pmod2$ T$^\dagger$ gates, and the total parity of T$^\dagger$ gates fixes whether the check stabilizes $T^\dagger|{+}\rangle$ or $T|{+}\rangle$. The wrong parity halves acceptance, breaks Definition~\ref{def:dc} and drops the fault distance to two, while at $\delta=3$ transversal T conjugates logical X like logical T$^\dagger$. Second, random outcomes during growth leave X-type Pauli frames that must be undone before the $\delta=5$ T layers~\cite{soft}. The following proposition relates the growth branches.

\begin{proposition}\label{prop:frames}
Let a correct non-adaptive circuit project onto $(I+sQ)/2$ at a point where a Pauli operator $G$ that anticommutes with $Q$ stabilizes every state that the fault-free skeleton produces there from its inputs. Let a second circuit agree with the first, except that it projects onto $(I-sQ)/2$ and then applies $G$ ideally. For every fault configuration $F$, the Kraus operators of the two circuits then agree up to a sign. The two circuits therefore reject the same configurations and have the same harmful configurations, fault distance and leading term. If $G$, carried through the Clifford operations that follow the projection, is applied instead by single-qubit gates at later points where the first circuit has idle locations, the two circuits have the same possible faults and the same fault distance, and also the same leading term if the gates and the idle locations carry the same noise.
\end{proposition}

\begin{proof}
In the skeleton of Theorem~\ref{thm:main}, the pre-projection Clifford map $M_{\mathrm{pre}}$ satisfies $GM_{\mathrm{pre}}=M_{\mathrm{pre}}$ after fixing the sign of $G$. Faults before the projection turn it into $K=\zeta PA_\sigma M_{\mathrm{pre}}B_\sigma^{\mathsf T}$, which $G$ maps to $\pm K$. Since $\{G,Q\}=0$, $(I-sQ)/2=G(I+sQ)G/2$, giving $G(I-sQ)K=\pm(I+sQ)K$, so the two Kraus operators agree up to sign. Single-qubit gates replacing idle locations carry the same Pauli fault set, and the same weights when gate and idle noise match.
\end{proof}

Stabilizer tracking verifies that every growth outcome satisfies Proposition~\ref{prop:frames} with X corrections at idle locations before the T layers. Every corrected branch therefore shares the fault distance $3$ and uniform-noise leading term $1.5\,p^3$ ($9$ malignant class triples of two hooks and one data fault, relative acceptance $2^{-4}$, bad weight $2^{-5}$ or $2^{-4}$, Table~\ref{tab:full}), extending Refs.~\cite{chan,wanzapirain}. Its S proxy has fault distance $4$ ($5.3\,p^4$). Table~\ref{tab:rates} compares sampled real-T and proxy rates under uniform and SI1000~\cite{si1000} noise with Ref.~\cite{soft}. At $p=10^{-3}$ under uniform noise our $3.26(21)\times10^{-9}$ (acceptance $0.144$) and proxy ratio $9.4(1.2)$ agree with the exact $3.33\times10^{-9}$ and ratio $9.6$ of Ref.~\cite{wanzapirain} on the circuit of Refs.~\cite{soft,clifft}.

\begin{table*}[t]
\caption{Logical error per accepted output of the circuits of Ref.~\cite{cultivation} with real T gates and with the S proxy (equal acceptance), alongside the GPU simulation of a modified $\delta=5$ circuit in Ref.~\cite{soft}. At $\delta=5$ the proxy error, of order $p^4$, is sampled directly with stim.}
\label{tab:rates}
\begin{ruledtabular}
\begin{tabular}{llcccc}
circuit & noise, $p$ & acceptance & real T & proxy & Ref.~\cite{soft} \\
\hline
$\delta=3$ & SI1000, $10^{-3}$ & 0.723 & $8.00(21)\times10^{-7}$ & $2.84(22)\times10^{-7}$ & \\
 & SI1000, $5\times10^{-4}$ & 0.850 & $1.52(4)\times10^{-7}$ & $3.35(35)\times10^{-8}$ & \\
$\delta=5$ & uniform, $2\times10^{-3}$ & 0.021 & $3.6(7)\times10^{-8}$ & $1.15(14)\times10^{-8}$ & $3.4\times10^{-8}$ \\
 & uniform, $10^{-3}$ & 0.144 & $3.26(21)\times10^{-9}$ & $3.47(40)\times10^{-10}$ & $4.6\times10^{-9}$ \\
 & uniform, $5\times10^{-4}$ & 0.379 & $2.72(12)\times10^{-10}$ & & $1.6\times10^{-10}$ \\
 & SI1000, $10^{-3}$ & 0.155 & $2.2(6)\times10^{-9}$ & $3.78(40)\times10^{-10}$ & \\
 & SI1000, $5\times10^{-4}$ & 0.394 & $1.09(8)\times10^{-10}$ & & \\
\end{tabular}
\end{ruledtabular}
\end{table*}

\subsection{Surface-code cultivation on \texorpdfstring{$\mathbb{RP}^2$}{RP2}}

We use the published circuits of Ref.~\cite{chenrp2} under uniform noise, with their final stabilizer measurements as ideal decoder. The $\delta=3$ circuit injects into a nine-qubit $\mathbb{RP}^2$ code, expands to $[[15,1,3]]$, and checks $X^{\otimes15}$ with $31$ T/T$^\dagger$ gates and a flagged CNOT contraction tree. Across all four first-round branches it has real-T fault distance $2$ ($12$ malignant pairs of a contraction-tree CNOT X hook and a data fault, relative acceptance $2^{-2}$ or $2^{-3}$, leading term $0.29\,p^2$) against proxy distance $3$. The $\delta=5$ circuit grows to $\mathbb{RP}^2$ at distance five and checks a self-dual $[[35,1,5]]$ code with $\pm1$ Z stabilizers ($101$ T gates, $\kappa=27$). Across all $256$ branches of its eight random outcomes it has real-T fault distance $3$ ($4$ malignant triples combining a growth Z fault with two faults on one data qubit of the last check, leading term $0.77\,p^3$) against proxy distance $4$.

\subsection{The circuit run on hardware}\label{sm:google}

Reference~\cite{googlecultivation} ran a $\delta=3$ color code on a square CZ grid ($14$ qubits, $7$ data) using injection, time-reversed GHZ-fold checks and error-correction cycles. In the published stim circuits we replace each tagged non-Clifford identity implementing $R_y(\pm\pi/4)$ by $H_{YZ}$ conjugating a noiseless virtual T or T$^\dagger$ gate, choose the injection sign for deterministic fault-free acceptance, add SI1000 noise, and cut the Fig.~3 circuit after the first, second or third check (the kickback measurement) or the following cycle before an ideal decoder.

\begin{proposition}[computer-assisted]\label{prop:google}
After one, two or three checks, the circuit of Ref.~\cite{googlecultivation} has real-T fault distance two, while its proxy fault distance is three. After two checks, each of its $20$ malignant pairs contains a fault in the unfolding of the last check, and an ideal injection leaves the same $20$ pairs.
\end{proposition}

Every cut has $20$ malignant pairs with leading terms from $0.44\,p^2$ to $2.2\,p^2$. For the two-check circuit, acceptance is $22\%$ at the experimental noise $p=2.3\times10^{-3}$, with logical error $2.4\times10^{-5}$ for real T against $6.9\times10^{-6}$ for the proxy, and the ratio grows from $3.1$ at $p=3\times10^{-3}$ to $4.5$ at $1.5\times10^{-3}$ and $6.1$ at $10^{-3}$. Note that the $0.9999(1)$ kickback fidelity at $8\%$ retention reported in Ref.~\cite{googlecultivation} also includes kickback-readout and non-Pauli errors, whereas our numbers evaluate the cultivated state prior to readout.

\begin{table*}[t]
\caption{Proxy and real-T fault distances, malignant configuration counts (fault-class tuples with relative acceptance $2^{-(\delta-1)}$ in the first two rows), and leading error per accepted output under uniform noise $p$ with idling (SI1000 in the last row, Proposition~\ref{prop:google}). Distances up to $\delta=7$ are certified by exact searches (Theorems~\ref{thm:main} and~\ref{thm:gauss}) and at $\delta=9$ by Proposition~\ref{prop:singleton} and floating-point MILP infeasibility (Appendix~\ref{sec:certificates}, Gurobi and HiGHS for the ladder, Gurobi for the verified cat). Unless marked ``no rounds'' or ``NN'' (nearest-neighbor, reversed cat growth, one check round, no stabilizer rounds), checks start from an ideal input between flagged stabilizer rounds, while the $\delta=3$ verified cat and NN protocol include noisy state preparation.}
\label{tab:full}
\begin{ruledtabular}
\begin{tabular}{lccccc}
check & $\delta$ & proxy & real T & malignant & leading term \\
\hline
GHZ fold~\cite{cultivation} & 3 & 3 & 2 & 9 & $0.43\,p^2$ \\
GHZ fold~\cite{cultivation} & 5 & 4 & 3 & 9 & $1.5\,p^3$ \\
GHZ fold, 6 trees & 7 & 7 & \textbf{4} & 15 to 957 & $(34$ to $4.4\times10^3)\,p^4$ \\
cat, verified & 3 & 3 & \textbf{3} & 2405 & $4.2\times10^3p^3$ \\
cat, not verified & 5 & 5 & 3 & 99 & $0.62\,p^3$ \\
cat, verified & 5 & 5 & \textbf{5} & 754{,}515 & $1.3\times10^6p^5$ \\
cat, verified, 2 rounds & 5 & 5 & \textbf{5} & & \\
cat, not verified & 7 & 7 & 4 & 120 & $0.12\,p^4$ \\
cat, verified & 7 & 7 & \textbf{7} & & \\
cat, verified, no rounds & 7 & 7 & \textbf{7} & & \\
cat, verified, no rounds & 9 & 9 & \textbf{9} & & \\
NN cat, ring & 3 & 3 & \textbf{3} & 574 & $6.7\times10^2p^3$ \\
NN cat, ladder & 5 & 5 & \textbf{5} & 85{,}464 & $5.3\times10^6p^5$ \\
NN cat, ladder & 7 & 7 & \textbf{7} & & \\
NN cat, ladder & 9 & 9 & \textbf{9} & & \\
NN protocol & 3 & 3 & \textbf{3} & 10{,}573 & $2.8\times10^3p^3$ \\
Google experiment & 3 & 3 & 2 & 20 & $2.2\,p^2$ \\
\end{tabular}
\end{ruledtabular}
\end{table*}

\section{Checks that keep the full distance}\label{sm:nn}

This appendix gives the checks of Sec.~\ref{sec:design}: Shor-style checks at every distance, the verified cats and nearest-neighbor checks that we certified, and their costs.

\subsection{Shor-style checks at every distance}\label{sec:shor}

\begin{corollary}\label{cor:shor}
Let a stabilizer code of distance $\delta$ have a check $UX^{\otimes n}U^\dagger$ with a layer $U$ of single-qubit gates, whose outcome on the encoded target is deterministic, as for the triangular color codes with T and T$^\dagger$ gates. Consider a circuit that starts from the ideally encoded target, measures the check once or several times, each time by the layer $U^\dagger$, a measurement of $X^{\otimes n}$ and the layer $U$, possibly between stabilizer measurements, and ends in the ideal decoder. If every measurement uses the gadget of Lemma~\ref{lem:cat} with $\theta=\delta-1$ rounds, the circuit has real-T fault distance $\delta$ if its data can fail just before the decoder. With the cat grown from the middle of a path of grid sites, one next to each data qubit, and verified by the parities of neighboring sites, all two-qubit gates of the check act between neighbors of a square grid with $3n$ qubits.
\end{corollary}

\begin{proof}
With fewer than $\delta$ faults in total, faults outside gadgets act on one data qubit each and each gadget has at most $\theta=\delta-1$ faults, leaving $R$ or $Rg$ on at most that many qubits by Lemma~\ref{lem:cat}, so Theorem~\ref{thm:strong} and Proposition~\ref{prop:singleton} give $d=\delta$. Placing data qubit $q$ at $(q,-1)$, its cat site at $(q,0)$ and an ancilla at $(q,1)$, each parity $Z_{(x,0)}Z_{(x+1,0)}$ is measured by resetting $(x,1)$ and $(x+1,1)$, applying $\mathrm{CX}_{(x,0),(x,1)}$, $\mathrm{CX}_{(x+1,0),(x+1,1)}$, $\mathrm{CX}_{(x+1,1),(x,1)}$ and $\mathrm{CX}_{(x+1,0),(x+1,1)}$, and measuring both ancillas. Cat qubits act only as controls along a connected path, so Lemma~\ref{lem:cat} applies.
\end{proof}

Exact evaluation confirms this $3n$-qubit layout at $\delta=3$ and $5$ ($21$ and $57$ qubits, $\theta=2$ and $4$), while Proposition~\ref{prop:nn} reduces verification to a single round up to $\delta=9$.

\subsection{Verified cats}\label{sec:vcat}

At $\delta=5$ and $7$ the all-to-all verified-cat checks start from an ideally encoded input ($T|{+}\rangle$ at $\delta=5$, $T^\dagger|{+}\rangle$ at $\delta=7$) and place one Shor-style check round~\cite{shor}, or two separated by a stabilizer round, between flagged stabilizer rounds~\cite{chaoreichardt}, under uniform noise on gates, resets, measurements and idle qubits and with a noiseless final decoder. At $\delta=3$, a noisy injection of $T^\dagger|{+}\rangle$ and a CNOT encoder precede two check rounds and three flagged rounds. Each check round applies $U^\dagger$, grows an $n$-qubit cat by a tree of CNOTs from $|{+}\rangle$, verifies all $n-1$ parities $Z_{c_q}Z_{c_{q+1}}$ on fresh ancillas, applies $\mathrm{CX}_{c_q,q}$ to the data, measures the cat in X and applies $U$. Exhaustive searches find no harmful configuration of four faults at $\delta=5$ and none of up to six faults at $\delta=7$, whereas without verification harmful configurations appear at three and four faults (Table~\ref{tab:full}). Without stabilizer rounds the verified cat keeps $d=7$ at $\delta=7$ ($110$ qubits) and $d=9$ at $\delta=9$ ($182$ qubits, Appendix~\ref{sec:certificates}). The price is acceptance. Under uniform noise at $p=10^{-3}$ and without gathered schedules, going from $\delta=5$ to $7$ lowers it from $5\%$ to $0.006\%$ for the verified cat, from $15\%$ to $0.4\%$ for the unverified cat and from $1.7\%$ to $0.002\%$ for the GHZ fold.

\begin{table*}[t]
\caption{Check costs under SI1000 noise at $p=10^{-3}$, showing qubits, circuit ticks (including one noiseless encoder tick), fault-free-relative acceptance, real-T fault distance (FD), leading wrong-output probability, postselected logical error $\epsilon$, gathered-schedule error $\epsilon_{\rm g}$ (Appendix~\ref{sec:gathered}, same FD), and spacetime volume ($\text{qubits}\times\text{layers}/\text{acc.}$). Rows marked $^*$ include noisy state preparation, while other rows start from an ideal input (between two flagged stabilizer rounds unless marked ``no rounds'' or ``NN'', with GHZ folds taking the lowest-coefficient tree). Values of $\epsilon$ sum up to eight fault classes via Appendix~\ref{sm:algorithm} (eight-class share $<0.3\%$, sampled directly for Refs.~\cite{cultivation,googlecultivation}), with one standard error in parentheses when $\geq2\%$, exact weight-$\leq5$ lower bounds when $\text{acc.}<5\%$, and $<3\times10^{-7}$ indicating zero accepted runs in $10^7$ shots.}
\label{tab:cost}
\begin{ruledtabular}
\begin{tabular}{lccccccccc}
check & $\delta$ & qubits & layers & acc. & FD & leading term & $\epsilon$ & $\epsilon_{\rm g}$ & volume \\
\hline
Ref.~\cite{cultivation}$^*$ & 3 & 15 & 39 & 0.72 & 2 & & $8.0(2)\times10^{-7}$ &  & $8.1\times10^2$ \\
Google$^*$, NN & 3 & 14 & 156 & 0.52 & 2 & $2.2\,p^2$ & $3.1\times10^{-6}$ &  & $4.2\times10^3$ \\
GHZ fold & 3 & 38 & 46 & 0.34 & 2 & $0.52\,p^2$ & $4.5\times10^{-6}$ &  & $5.2\times10^3$ \\
GHZ fold, no rounds & 3 & 14 & 19 & 0.81 & 2 & $0.29\,p^2$ & $7.5\times10^{-7}$ & $2.3\times10^{-7}$ & $3.3\times10^2$ \\
cat, verified$^*$ & 3 & 69 & 34 & 0.21 & \textbf{3} & $1.7\times10^4p^3$ & $2.0\times10^{-5}$ &  & $1.1\times10^4$ \\
cat, verified, no rounds & 3 & 20 & 19 & 0.78 & \textbf{3} & $6.8\times10^2p^3$ & $7.4\times10^{-7}$ & $3.1\times10^{-8}$ & $4.9\times10^2$ \\
NN ring, direct X readout & 3 & 17 & 10 & 0.86 & \textbf{3} & $3.4\times10^2p^3$ & $3.6\times10^{-7}$ & $7.1\times10^{-8}$ & $2.0\times10^2$ \\
NN ring, growth reversed & 3 & 17 & 15 & 0.80 & \textbf{3} & $1.2\times10^3p^3$ & $1.3\times10^{-6}$ & $1.6\times10^{-8}$ & $3.2\times10^2$ \\
NN protocol$^*$ & 3 & 17 & 107 & 0.11 & \textbf{3} & $7.8\times10^3p^3$ & $1.45(4)\times10^{-5}$ & $4.70(15)\times10^{-6}$ & $1.7\times10^4$ \\
\hline
Ref.~\cite{cultivation}$^*$ & 5 & 42 & 94 & 0.16 & 3 & & $2.2(6)\times10^{-9}$ &  & $2.5\times10^4$ \\
GHZ fold & 5 & 110 & 74 & 0.0047 & 3 & $0.34\,p^3$ & $\geq6.7\times10^{-8}$ &  & $1.7\times10^6$ \\
GHZ fold, no rounds & 5 & 38 & 21 & 0.56 & 3 & $0.035\,p^3$ & $4.36(19)\times10^{-10}$ & $2.52(5)\times10^{-10}$ & $1.4\times10^3$ \\
cat, not verified & 5 & 110 & 36 & 0.047 & 3 & $0.66\,p^3$ & $\geq4.5\times10^{-9}$ &  & $8.4\times10^4$ \\
cat, not verified, no rounds & 5 & 38 & 9 & 0.82 & 3 & $0.091\,p^3$ & $1.27\times10^{-10}$ & $1.63(5)\times10^{-10}$ & $4.2\times10^2$ \\
cat, verified & 5 & 128 & 51 & 0.0097 & \textbf{5} & $1.5\times10^7p^5$ & $\geq1.5\times10^{-8}$ &  & $6.7\times10^5$ \\
cat, verified, no rounds & 5 & 56 & 45 & 0.21 & \textbf{5} & $1.4\times10^7p^5$ & $1.93(16)\times10^{-8}$ & $3.7(1.2)\times10^{-12}$ & $1.2\times10^4$ \\
NN ladder, direct X readout & 5 & 46 & 24 & 0.29 & \textbf{5} & $8.4\times10^6p^5$ & $9.6(6)\times10^{-9}$ & $7.7(1.1)\times10^{-11}$ & $3.8\times10^3$ \\
NN ladder, growth reversed & 5 & 46 & 34 & 0.15 & \textbf{5} & $5.5\times10^7p^5$ & $7.2(5)\times10^{-8}$ & $8.6(1.0)\times10^{-11}$ & $1.1\times10^4$ \\
\hline
GHZ fold & 7 & 218 & 114 & $<3\times10^{-7}$ & 4 & $1.6\times10^2p^4$ &  & & $>8\times10^{10}$ \\
cat, not verified & 7 & 218 & 51 & $8.6\times10^{-5}$ & 4 & $0.17\,p^4$ &  & & $1.3\times10^8$ \\
cat, verified & 7 & 254 & 88 & $<3\times10^{-7}$ & \textbf{7} & & &  & $>7\times10^{10}$ \\
cat, verified, no rounds & 7 & 110 & 82 & 0.0045 & \textbf{7} & & &  & $2.0\times10^6$ \\
NN ladder, direct X readout & 7 & 91 & 33 & 0.030 & 6 & & &  & $1.0\times10^5$ \\
NN ladder, growth reversed & 7 & 91 & 52 & 0.0028 & \textbf{7} & & &  & $1.7\times10^6$ \\
\end{tabular}
\end{ruledtabular}
\end{table*}

\subsection{Nearest-neighbor checks and gathered schedules}\label{sec:nncheck}\label{sec:gathered}

The square-grid checks of Proposition~\ref{prop:nn} (Fig.~\ref{fig:nn}) start from an ideally encoded state without stabilizer rounds, and their operations are scheduled as early as possible, with each CNOT counted as one two-qubit gate. Each check applies $U^\dagger$, grows a path cat from its middle, verifies it once, couples every data qubit to its adjacent path site by a CNOT, un-grows the cat onto its root and applies $U$. The root is measured in X, while the other path sites return to $|0\rangle$ and are measured in Z as detectors. The $\delta=3$ ring has nine path sites, seven next to data and two bridging corners, verified by the parity $Z_aZ_b$ of its ends. The $\delta=5$ ladder has $19$ path sites on the rows $(x,0)$ and $(x,2)$, $x=0,\dots,7$, joined by the turn $(8,0),(8,1),(8,2)$ with root $(8,1)$. The eight middle sites $(x,1)$ measure the rungs $Z_{(x,0)}Z_{(x,2)}$, each on one ancilla, and the seven diagonals $Z_{(x,0)}Z_{(x-1,2)}$, each on the adjacent ancillas $a_1=(x,1)$ and $a_2=(x-1,1)$ through $\mathrm{CX}_{(x,0),a_1}$, $\mathrm{CX}_{(x-1,2),a_2}$, $\mathrm{CX}_{a_2,a_1}$ and $\mathrm{CX}_{(x-1,2),a_2}$, so that path qubits act only as controls. Rungs catch X spread along one arm and diagonals catch mirror spreads on both arms. Without the diagonals the fault distance drops to four. The ladders at $\delta=7$ and $9$ have $37$ and $61$ path sites and $17$ and $29$ middle ancillas ($91$ and $151$ qubits). At $\delta=7$ an exhaustive search finds no harmful configuration of up to six faults, and the proxy also has fault distance seven. At $\delta=9$ the ladder is certified by Appendix~\ref{sec:certificates}.

Reading the cat by direct X measurements instead of un-growing it shortens the idling of the data at $\delta=3$ and $5$ (Table~\ref{tab:cost}), but lowers the fault distance to six at $\delta=7$. SI1000 noise charges $2p$ to qubits that idle during a measurement or reset, so the ungathered full-distance checks at $\delta=5$ are limited by the idling of the data above $p\approx10^{-4}$. Gathering removes most of this cost. Delaying resets and measurements by up to eight layers so that they share layers, and deferring the final measurements to one last layer, leaves $2$ of $15$ layers with measurements or resets in the reversed $\delta=3$ ring and $7$ of $26$ in the direct $\delta=5$ ladder, and it preserves every fault distance. At $p=10^{-3}$ it raises the acceptance to $89\%$ for the ring, $43\%$ for the reversed $\delta=5$ ladder and $58\%$ for the $\delta=5$ verified cat without rounds, and from $0.45\%$ to $25\%$ for the $\delta=7$ verified cat without rounds. With gathered schedules, every full-distance check from an ideal input outperforms every half-distance check at $\delta=3$ and $5$ (Table~\ref{tab:cost}).

\subsection{A complete protocol}\label{sec:complete}

To prepare the state on the $\delta=3$ grid ($17$ qubits) with all operations noisy, we apply $T^\dagger$ to one $|{+}\rangle$ data qubit, relay a CNOT encoder along the cat path, measure all six stabilizers via verified path cats to catch encoder hooks, and apply three ring checks (real-T and proxy $d=3$, dropping to $d=2$ without the X stabilizers). Across $107$ layers under SI1000 noise, gathering improves the leading term and $p=10^{-3}$ performance from $7.8\times10^3p^3$ ($11\%$ acceptance, $\epsilon=1.45\times10^{-5}$) to $2.4\times10^3p^3$ ($22\%$ acceptance, $\epsilon=4.7\times10^{-6}$), showing that stabilizer measurements around the checks dominate the cost.

\bibliography{cultivation}

@misc{distillationpaper,
  author        = {Balada Gaggioli, Lloren\c{c} and Mare\v{c}ek, Jakub},
  title         = {The limits of magic-state distillation},
  year          = {2026},
  eprint        = {2610.31115},
  archivePrefix = {arXiv},
  primaryClass  = {quant-ph}
}

@article{stim,
  author  = {Gidney, Craig},
  title   = {Stim: a fast stabilizer circuit simulator},
  journal = {Quantum},
  volume  = {5},
  pages   = {497},
  year    = {2021},
  doi     = {10.22331/q-2021-07-06-497}
}

@misc{delfossepaetznick,
  author        = {Delfosse, Nicolas and Paetznick, Adam},
  title         = {Spacetime codes of {Clifford} circuits},
  year          = {2023},
  eprint        = {2304.05943},
  archivePrefix = {arXiv},
  primaryClass  = {quant-ph}
}

@article{bombinzx,
  author  = {Bomb\'in, H\'ector and Litinski, Daniel and Nickerson, Naomi and Pastawski, Fernando and Roberts, Sam},
  title   = {Unifying flavors of fault tolerance with the {ZX} calculus},
  journal = {Quantum},
  volume  = {8},
  pages   = {1379},
  year    = {2024},
  doi     = {10.22331/q-2024-06-18-1379}
}

@article{li2015,
  author  = {Li, Ying},
  title   = {A magic state's fidelity can be superior to the operations that created it},
  journal = {New J. Phys.},
  volume  = {17},
  pages   = {023037},
  year    = {2015},
  doi     = {10.1088/1367-2630/17/2/023037}
}

@article{chamberlandnoh,
  author  = {Chamberland, Christopher and Noh, Kyungjoo},
  title   = {Very low overhead fault-tolerant magic state preparation using redundant ancilla encoding and flag qubits},
  journal = {npj Quantum Inf.},
  volume  = {6},
  pages   = {91},
  year    = {2020},
  doi     = {10.1038/s41534-020-00319-5}
}

@article{itogawa,
  author  = {Itogawa, Tomohiro and Takada, Yugo and Hirano, Yutaka and Fujii, Keisuke},
  title   = {Efficient magic state distillation by zero-level distillation},
  journal = {PRX Quantum},
  volume  = {6},
  pages   = {020356},
  year    = {2025},
  doi     = {10.1103/thxx-njr6}
}

@article{knilllaflamme,
  author  = {Knill, Emanuel and Laflamme, Raymond},
  title   = {Theory of quantum error-correcting codes},
  journal = {Phys. Rev. A},
  volume  = {55},
  pages   = {900--911},
  year    = {1997},
  doi     = {10.1103/PhysRevA.55.900}
}

@article{aaronsongottesman,
  author  = {Aaronson, Scott and Gottesman, Daniel},
  title   = {Improved simulation of stabilizer circuits},
  journal = {Phys. Rev. A},
  volume  = {70},
  pages   = {052328},
  year    = {2004},
  doi     = {10.1103/PhysRevA.70.052328}
}

@article{zhouleungchuang,
  author  = {Zhou, Xinlan and Leung, Debbie W. and Chuang, Isaac L.},
  title   = {Methodology for quantum logic gate construction},
  journal = {Phys. Rev. A},
  volume  = {62},
  pages   = {052316},
  year    = {2000},
  doi     = {10.1103/PhysRevA.62.052316}
}

@article{chaoreichardt,
  author  = {Chao, Rui and Reichardt, Ben W.},
  title   = {Quantum error correction with only two extra qubits},
  journal = {Phys. Rev. Lett.},
  volume  = {121},
  pages   = {050502},
  year    = {2018},
  doi     = {10.1103/PhysRevLett.121.050502}
}

@misc{gsjcode,
  author       = {Gidney, Craig and Shutty, Noah and Jones, Cody},
  title        = {Code for ``{M}agic state cultivation: growing {T} states as cheap as {CNOT} gates''},
  howpublished = {\url{https://github.com/Strilanc/magic-state-cultivation}, commit 871e68f},
  year         = {2024}
}

@article{chamberlandcross,
  author  = {Chamberland, Christopher and Cross, Andrew W.},
  title   = {Fault-tolerant magic state preparation with flag qubits},
  journal = {Quantum},
  volume  = {3},
  pages   = {143},
  year    = {2019},
  doi     = {10.22331/q-2019-05-20-143}
}

@article{bravyigosset,
  author  = {Bravyi, Sergey and Gosset, David},
  title   = {Improved classical simulation of quantum circuits dominated by {Clifford} gates},
  journal = {Phys. Rev. Lett.},
  volume  = {116},
  pages   = {250501},
  year    = {2016},
  doi     = {10.1103/PhysRevLett.116.250501}
}

@article{chenrp2,
  author  = {Chen, Zi-Han and Chen, Ming-Cheng and Lu, Chao-Yang and Pan, Jian-Wei},
  title   = {Efficient magic state cultivation on {$\mathbb{RP}^2$}},
  journal = {PRX Quantum},
  volume  = {7},
  pages   = {010315},
  year    = {2026},
  doi     = {10.1103/9kys-3whh}
}

@article{sahay,
  author  = {Sahay, Kaavya and Tsai, Pei-Kai and Chang, Kathleen and Su, Qile and Smith, Thomas B. and Singh, Shraddha and Puri, Shruti},
  title   = {Fold-transversal surface code cultivation},
  journal = {PRX Quantum},
  volume  = {7},
  pages   = {033006},
  year    = {2026},
  doi     = {10.1103/gpvl-lg4c}
}

@article{surti,
  author  = {Surti, Samyak and Daguerre, Lucas and Kim, Isaac H.},
  title   = {Efficient simulation of logical magic state preparation protocols},
  journal = {PRX Quantum},
  volume  = {7},
  pages   = {020329},
  year    = {2026},
  doi     = {10.1103/fby6-xjbm}
}

@article{wanzhong,
  author  = {Wan, Kwok Ho and Zhong, Zhenghao and Zapirain, Ainhoa},
  title   = {Simulating magic state cultivation with few {Clifford} terms},
  journal = {Quantum},
  volume  = {10},
  pages   = {2134},
  year    = {2026},
  doi     = {10.22331/q-2026-06-12-2134}
}

@misc{soft,
  author        = {Li, Riling and Zheng, Keli and Zhang, Yiming and Lou, Huazhe and Ying, Shenggang and Liu, Ke and Sun, Xiaoming},
  title         = {{SOFT}: a high-performance simulator for universal fault-tolerant quantum circuits},
  year          = {2025},
  eprint        = {2512.23037},
  archivePrefix = {arXiv},
  primaryClass  = {quant-ph}
}

@misc{clifft,
  author        = {Chase, Bradley A. and Labib, Farrokh},
  title         = {Clifft: fast exact simulation of near-{Clifford} quantum circuits},
  year          = {2026},
  eprint        = {2604.27058},
  archivePrefix = {arXiv},
  primaryClass  = {quant-ph}
}

@misc{symft,
  author        = {Fang, Wang and Lou, Huazhe and Li, Riling},
  title         = {{SymFT}: universal fault-tolerant quantum circuit simulation via symbolic {Clifford}--{Pauli} frames and stabilizer coordinates},
  year          = {2026},
  eprint        = {2607.28600},
  archivePrefix = {arXiv},
  primaryClass  = {quant-ph}
}

@misc{syqma,
  author        = {Umbrarescu, George and Amaro, David},
  title         = {{SyQMA}: a memory-efficient, symbolic and exact universal simulator for quantum error correction},
  year          = {2026},
  eprint        = {2604.15043},
  archivePrefix = {arXiv},
  primaryClass  = {quant-ph}
}

@misc{takada,
  author        = {Takada, Yugo and Bartlett, Stephen D. and Williamson, Dominic J.},
  title         = {Exact efficient simulation of noisy logical magic states using {Clifford} stabilizers},
  year          = {2026},
  eprint        = {2609.16929},
  archivePrefix = {arXiv},
  primaryClass  = {quant-ph}
}

@misc{chan,
  author        = {Chan, Tim and Strikis, Armands and Sun, Zhu and Cai, Zhenyu},
  title         = {Diagnosing and restoring the degraded fault distance of magic state cultivation},
  year          = {2026},
  eprint        = {2609.17706},
  archivePrefix = {arXiv},
  primaryClass  = {quant-ph}
}

@misc{wanzapirain,
  author        = {Wan, Kwok Ho and Zapirain, Ainhoa},
  title         = {Exact logical error rates for magic state cultivation},
  year          = {2026},
  eprint        = {2609.18922},
  archivePrefix = {arXiv},
  primaryClass  = {quant-ph}
}

@inproceedings{shor,
  author    = {Shor, Peter W.},
  title     = {Fault-tolerant quantum computation},
  booktitle = {Proceedings of the 37th Annual Symposium on Foundations of Computer Science},
  pages     = {56--65},
  publisher = {IEEE},
  year      = {1996},
  doi       = {10.1109/SFCS.1996.548464}
}

@article{rains,
  author  = {Rains, Eric M.},
  title   = {Quantum shadow enumerators},
  journal = {IEEE Trans. Inf. Theory},
  volume  = {45},
  pages   = {2361--2366},
  number  = {7},
  year    = {1999},
  doi       = {10.1109/18.796376}
}

@article{si1000,
  author  = {Gidney, Craig and Newman, Michael and McEwen, Matt},
  title   = {Benchmarking the planar honeycomb code},
  journal = {Quantum},
  volume  = {6},
  pages   = {813},
  year    = {2022},
  doi     = {10.22331/q-2022-09-21-813}
}

@misc{merlin,
  author        = {Cercelescu, Serban and Koch, Mark and Pesah, Arthur},
  title         = {Polynomial-time simulation of non-{Clifford} quantum error correction},
  year          = {2026},
  eprint        = {2610.06811},
  archivePrefix = {arXiv},
  primaryClass  = {quant-ph}
}

@misc{magna,
  author        = {Wang, Yifei and Gu, Yingfei},
  title         = {Magna: simulating magic state factories as cheaply as {Clifford} circuits},
  year          = {2026},
  eprint        = {2610.22315},
  archivePrefix = {arXiv},
  primaryClass  = {quant-ph}
}

@misc{spidercat,
  author        = {Khesin, Andrey Boris and Li, Sarah Meng and Po\'or, Boldizs\'ar and Rodatz, Benjamin and van de Wetering, John and Yeung, Richie},
  title         = {{SpiderCat}: optimal fault-tolerant cat state preparation},
  year          = {2026},
  eprint        = {2603.05391},
  archivePrefix = {arXiv},
  primaryClass  = {quant-ph}
}

@misc{kalraprakash,
  author        = {Kalra, Amolak Ratan and Prakash, Shiroman},
  title         = {Invariant theory, magic state distillation, and bounds on classical codes},
  year          = {2025},
  eprint        = {2501.10163},
  archivePrefix = {arXiv},
  primaryClass  = {quant-ph}
}

@misc{mincs,
  author        = {Min, Gunsik and Heo, Jun},
  title         = {Direct cultivation of entangled $|{CS}\rangle$ magic states},
  year          = {2026},
  eprint        = {2609.09993},
  archivePrefix = {arXiv},
  primaryClass  = {quant-ph}
}

@misc{minmulti,
  author        = {Min, Gunsik and Heo, Jun},
  title         = {Coded {Clifford} measurements for multiqubit magic-state cultivation},
  year          = {2026},
  eprint        = {2609.09994},
  archivePrefix = {arXiv},
  primaryClass  = {quant-ph}
}

@misc{hines,
  author        = {Hines, Jordan and Ostrove, Corey and Rudinger, Kenneth and Seritan, Stefan and Young, Kevin and Blume-Kohout, Robin and Proctor, Timothy},
  title         = {Simulating quantum error correction beyond {Pauli} stochastic errors},
  year          = {2026},
  eprint        = {2603.18457},
  archivePrefix = {arXiv},
  primaryClass  = {quant-ph}
}

@misc{mehta,
  author        = {Mehta, Rohan and Menon, Varun and Zhou, Hengyun and Lukin, Mikhail D. and Bonilla Ataides, J. Pablo},
  title         = {The practical cost of magic state cultivation},
  year          = {2026},
  eprint        = {2610.06828},
  archivePrefix = {arXiv},
  primaryClass  = {quant-ph}
}

@misc{hartweg,
  author        = {Hartweg, Tom and Pi{\~n}eiro Orioli, Asier},
  title         = {Comparing magic state cultivation methods using matrix product states},
  year          = {2026},
  eprint        = {2609.19116},
  archivePrefix = {arXiv},
  primaryClass  = {quant-ph}
}

@article{hetenyi,
  author  = {Het{\'e}nyi, Bence and Brown, Benjamin J. and Williamson, Dominic J.},
  title   = {Constant-depth magic state cultivation with {Clifford} measurements by gauging},
  journal = {Phys. Rev. Lett.},
  volume  = {137},
  pages   = {140601},
  year    = {2026},
  doi     = {10.1103/f7fz-6zsy}
}

@misc{hirasaki,
  author        = {Hirasaki, Yuta and Lee, Jong Yeon},
  title         = {Logical {CCZ} state cultivation},
  year          = {2026},
  eprint        = {2610.20591},
  archivePrefix = {arXiv},
  primaryClass  = {quant-ph}
}

@misc{chensqrtt,
  author        = {Chen, I-Chi and da Silva Fonseca, Matheus and Sornborger, Andrew},
  title         = {Efficient magic state cultivation for $\sqrt{T}$ gates},
  year          = {2026},
  eprint        = {2606.10430},
  archivePrefix = {arXiv},
  primaryClass  = {quant-ph}
}

@misc{tuloup,
  author        = {Tuloup, Thomas and Ayral, Thomas},
  title         = {Computing logical error thresholds with the {Pauli} frame sparse representation},
  year          = {2026},
  eprint        = {2603.14670},
  archivePrefix = {arXiv},
  primaryClass  = {quant-ph}
}

@misc{simsek,
  author        = {Simsek, Selwyn and Macpherson, Callum and Self, Chris N. and Amaro, David},
  title         = {Fast verification of fault tolerant magic state preparation circuits with the {CNOT} + {T} gate set},
  year          = {2026},
  eprint        = {2611.04909},
  archivePrefix = {arXiv},
  primaryClass  = {quant-ph}
}

@misc{garner,
  author        = {Garner, Sean and Hoyt, Aaron and Kan, Shuwen and Stein, Samuel and Liu, Chenxu and Li, Ang},
  title         = {Early fault-tolerant magic state synthesis and spacetime efficient analog rotations with color code on a superconducting quantum processor},
  year          = {2026},
  eprint        = {2611.00404},
  archivePrefix = {arXiv},
  primaryClass  = {quant-ph}
}

@article{vaknin,
  author  = {Vaknin, Yotam and Jacoby, Shoham and Grimsmo, Arne and Retzker, Alex},
  title   = {High rate magic state cultivation on the surface code},
  journal = {PRX Quantum},
  volume  = {7},
  pages   = {010353},
  year    = {2026},
  doi     = {10.1103/p8tw-6kq9}
}

@article{bravyikitaev,
  author  = {Bravyi, Sergey and Kitaev, Alexei},
  title   = {Universal quantum computation with ideal {Clifford} gates and noisy ancillas},
  journal = {Phys. Rev. A},
  volume  = {71},
  pages   = {022316},
  year    = {2005},
  doi     = {10.1103/PhysRevA.71.022316}
}

@article{bravyihaah,
  author  = {Bravyi, Sergey and Haah, Jeongwan},
  title   = {Magic-state distillation with low overhead},
  journal = {Phys. Rev. A},
  volume  = {86},
  pages   = {052329},
  year    = {2012},
  doi     = {10.1103/PhysRevA.86.052329}
}

@misc{cultivation,
  author        = {Gidney, Craig and Shutty, Noah and Jones, Cody},
  title         = {Magic state cultivation: growing {T} states as cheap as {CNOT} gates},
  year          = {2024},
  eprint        = {2409.17595},
  archivePrefix = {arXiv},
  primaryClass  = {quant-ph}
}

@article{googlecultivation,
  author        = {Rosenfeld, E. and Gidney, C. and Roberts, G. and others},
  title         = {Magic state cultivation on a superconducting quantum processor},
  journal       = {Science},
  pages         = {eaee7961},
  year          = {2026},
  doi           = {10.1126/science.aee7961},
  eprint        = {2512.13908},
  archivePrefix = {arXiv}
}

@article{aharonovbenor,
  author  = {Aharonov, Dorit and Ben-Or, Michael},
  title   = {Fault-tolerant quantum computation with constant error rate},
  journal = {SIAM J. Comput.},
  volume  = {38},
  pages   = {1207--1282},
  year    = {2008},
  doi     = {10.1137/S0097539799359385}
}

@article{agp,
  author  = {Aliferis, Panos and Gottesman, Daniel and Preskill, John},
  title   = {Quantum accuracy threshold for concatenated distance-3 codes},
  journal = {Quantum Inf. Comput.},
  volume  = {6},
  pages   = {97--165},
  year    = {2006},
  doi     = {10.26421/QIC6.2-1}
}

\end{document}